\documentclass[aps, prb, reprint, onecolumn, showpacs, amsmath, amssymb]{revtex4-1}
\usepackage{graphicx}
\begin{document}

\title{Signatures of exchange correlations in the thermopower of quantum dots}
\author{Gabriel Billings}
\affiliation{Department of Physics, Stanford University, Stanford, California 94305}
\author{A. Douglas Stone}
\affiliation{Department of Applied Physics, Post Office Box 208284, Yale University, New Haven, Connecticut 06520}
\author{Y. Alhassid}
\affiliation{Center for Theoretical Physics, Sloane Physics Laboratory, Yale University, New Haven, Connecticut 06520}
\date{\today}
\begin{abstract}
We use a many-body rate-equation approach to calculate the thermopower of a quantum dot in the presence of an exchange interaction. At temperatures much smaller than the single-particle level spacing, the known quantum jumps (discontinuities) in the thermopower are split by the exchange interaction.  The origin and nature of the splitting are elucidated with a simple physical argument based on the nature of the intermediate excited state in the sequential tunneling approach.  We show that this splitting is sensitive to the number parity of electrons in the dot and the dot's ground-state spin.  These effects are suppressed when cotunneling dominates the electrical and thermal conductances. We calculate the thermopower in the presence of elastic cotunneling, and show that some signatures of exchange correlations should still be observed with current experimental methods. In particular, we propose a method to determine the strength of the exchange interaction from measurements of the thermopower.
\end{abstract}
\pacs{73.23.Hk, 72.20.Pa, 73.63.Kv, 73.40.Gk}
\maketitle

\section{Introduction}

Lateral quantum dots have been studied extensively both
experimentally~\cite{marcusreview} and theoretically.~\cite{alhassid2}  Such dots exhibit effects associated with both charging energy and a discrete single-particle spectrum, as captured by the constant interaction (CI) model. In dots with more than $\sim 50-100$ electrons, the irregular shape of the confining potential often leads to chaotic classical dynamics, and the fluctuations of single-particle energies and wave functions follow random matrix theory (RMT). The transport properties of such dots exhibit mesoscopic fluctuations as a function of external magnetic field and/or shape.  Consequently, the electrical conductance of a weakly coupled dot exhibits Coulomb Blockade oscillations (a charging energy effect) and peak-height fluctuations that are well-described by RMT.~\cite{JSA,folk,chang} However, the CI plus RMT model does not explain the statistics of the observed residual fluctuations in peak spacings, and it was recognized that electron-electron interactions beyond charging energy play an important role.~\cite{sivan96} A universal Hamiltonian was shown to properly describe the leading interaction terms in a chaotic quantum dot in the limit of large Thouless conductance.~\cite{univH,ABG02} These interaction terms include a spin-exchange term and a Cooper channel term. The latter is repulsive in quantum dots and can be ignored. However, inclusion of the ferromagnetic exchange interaction term was shown to provide a good agreement between the calculated and observed peak spacing and peak height statistics in quantum dots.~\cite{alhassid3, patel1,patel2}

The thermoelectric properties of quantum dots were studied by Beenakker and Staring~\cite{beenakker1} using the sequential tunneling approach in the framework of the CI model. Of particular interest is the thermopower of the dot, ${\cal S} =- \Delta V/\Delta T$, where $\Delta V$ is the voltage induced by a temperature difference $\Delta T$ across the dot under the condition of vanishing electrical current. The charging energy gives rise to large periodic sawtooth oscillations of the thermopower (see Fig.~1) of magnitude $e/2TC$ (where $T$ is the temperature and $C$ is the dot's capacitance). The thermopower vanishes at the charge degeneracy point where the conductance has a peak.  This behavior of the thermopower originates in the breaking of particle-hole symmetry. For sequential transport to occur in the ``valleys'' between conductance peaks, a thermal fluctuation to an unfavorable charge state must overcome the charging gap. A valley of the conductance corresponds to an equilibrium state of $N$ electrons. On one side of this valley, the $N-1$ electron states are closer in energy (to the $N$-electron states) and carry a hole current, while on the other side of the valley the $N+1$ electron states are closer in energy and carry a particle current.  In the center of the valley there is a (thermally rounded) discontinuity in the thermopower as it switches between hole and particle transport, leading to the large-scale sawtooth pattern shown in Fig.~1. The thermopower vanishes at the charge degeneracy point where particle and hole transport are equally likely.

\begin{figure}
\includegraphics{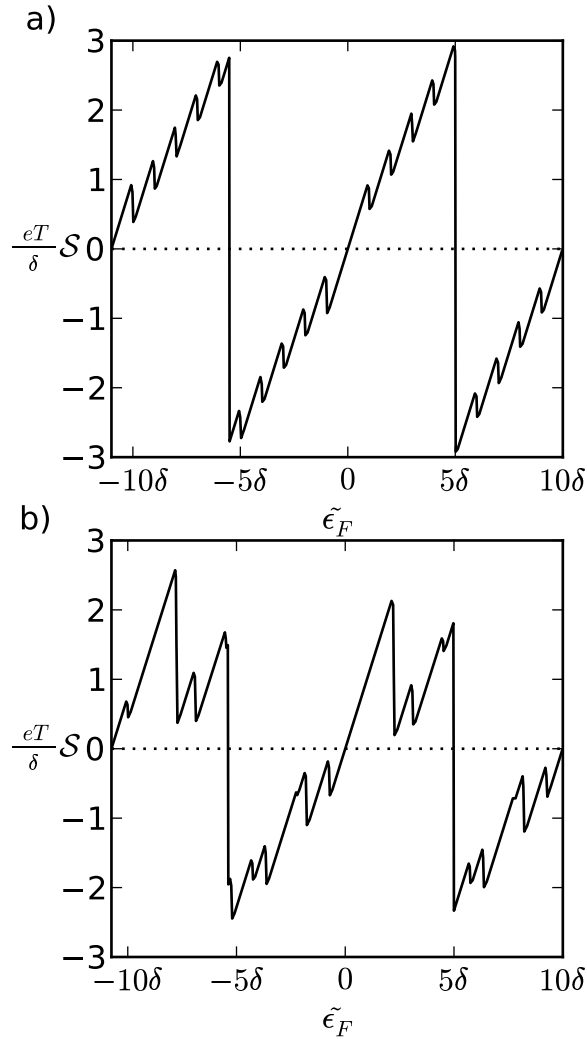}
\caption{Low-temperature limit of the thermopower ${\cal S}$ (in units of $\delta/eT$) versus the effective Fermi energy $\tilde \epsilon_F$ in the absence of exchange correlations ($\tilde \epsilon_F$ is controlled by a gate voltage). The dot has 20 electrons with $e^{2}/2C=5\delta$ and $kT=\delta/100$ ($\delta$ is the single-particle level spacing). In panel a) the level spacings and transition widths are uniform. In panel b) the level spacings and widths are drawn from the gaussian orthogonal ensemble of RMT. Here (and in all subsequent figures) $\tilde{\epsilon}_{F}$ is taken to be $0$ at the degeneracy point (where the conductance has a peak). The large scale jumps at the center of the conductance valleys (at $\pm 5\delta$) correspond to a change in the excited charge state that is closest to the dot's ground state.  Each sawtooth (e.g., between $-5\delta$ and $5\delta$) in both panels contains $ \sim 10$ fine structure (quantum) jumps corresponding to five accessible particle and five accessible hole single-particle levels in the dot.}
\end{figure}

Beyond the charging energy effect on the thermopower,  Ref.~\onlinecite{beenakker1} also predicted smaller ``teeth" structures superimposed on the larger scale sawtooth behavior at temperatures below the single-particle level spacing.  We will refer to these fine structure features as {\it quantum jumps} in the thermopower. This additional structure comes from the discrete nature of the manifold of excited states that contribute to the thermopower.  In the center of a valley, a large thermal fluctuation is required to transport charge through the dot. This fluctuation energy can be divided arbitrarily between the electron's states in the leads and states of the dot with equal probability, allowing a number of excited states in the dot to contribute to transport.  As the Fermi energy is varied away from the valley's center, the size of the thermal fluctuation necessary to allow transport through the dot decreases, thus reducing the number of excited states on the dot contributing to charge and energy transport. The effect is that the magnitude of the thermopower increases more rapidly than the classical prediction between sharp negative steps that occur as each excited level falls out of the accessible energy range. Thus, these steps in the thermopower locate (a subset) of the excited states on the dot at fixed number of electrons. We note that the example discussed in Ref.~\onlinecite{beenakker1} corresponds to an equally-spaced single-particle spectrum and equal level widths. Mesoscopic fluctuations have not been included in any of the theoretical studies of the quantum dot thermopower; these fluctuations cause the quantum jumps to deviate from uniform spacings and the size of these jumps to vary substantially.  In Fig.~1b we show a typical thermopower of the Beenakker-Staring theory in the presence of random matrix fluctuations in both spacings and widths, as compared with the case of an equally spaced single-particle spectrum and equal level widths in Fig.~1a.

Experiments by Staring {\it et al.}~\cite{staring}  and Dzurak {\it et al.}~\cite{dzurak1} found evidence of the large-scale classical sawtooth structure, but with peak-to-peak amplitude that is significantly smaller than the theoretical prediction.  In later work, Dzurak {\it et al.}~\cite{dzurak2}  found clear evidence at low temperatures of the quantum jumps predicted by the theory, but only near the degeneracy point and not in the center of the valleys.  Subsequently, this significant discrepancy between theory and experiment was explained by Turek and Matveev,~\cite{turek} who argued that cotunneling transport should dominate both the thermal and electrical conductances near the valley center and greatly reduce its magnitude compared to the predictions of the sequential tunneling theory. Furthermore, the thermopower lineshape is distorted from a pure sawtooth, reaching a maximum away from the valley center and then decreasing as cotunneling begins to dominate.  Effects of sequential tunneling and cotunneling on the thermopower were also studied in single molecules that are coupled to metallic leads.~\cite{koch04}

The Beenakker-Staring theory was formulated in the context of the CI model and did not include the effects of exchange correlations; hence spin entered trivially as a degeneracy of the single-particle levels. Here we are primarily concerned with novel effects that originate in the exchange interaction and appear in the quantum structure of the thermopower. The quantum structure is characteristic of sequential tunneling transport and not of cotunneling. The latter describes a coherent sum over many levels and smoothes out fine structure effects in the thermopower. The major part of this work is thus focused on the sequential tunneling thermopower, in which we find dramatic exchange interaction effects (in the absence of cotunneling). We then proceed to include cotunneling effects which impose a cut-off on the observability of these dramatic effects. This cut-off is rather stringent because the sequential tunneling in the conductance valleys is a thermally activated process, whereas cotunneling is not. Using current experimental methods and realistic device parameters, it is therefore difficult to see more than a single quantum jump in the vicinity of each Coulomb blockade peak. Nevertheless, we argue that the observability of {\it any} quantum jump is greatly enhanced by the exchange interaction effects, suggesting that exchange might have played a crucial role in the observation of quantum jumps by Dzurak {\it et al.}~\cite{dzurak2} Exploiting this sensitivity to exchange correlations, we propose a method to determine $J_s$ from an ensemble of measured thermopower traces.

Despite the constraints on the observability of the pure sequential tunneling thermopower, its fine structure remains of interest because, unlike the conductance, thermopower directly probes the excited states of the quantum dot in linear response (i.e., for a small source-drain voltage). Here we generalize the Beenakker-Staring theory, formulated in the CI model, to a dot with electron-electron interactions beyond charging energy and in particular to the universal Hamiltonian framework, allowing the study of exchange interaction effects. We  find that the quantum structure in the thermopower provides information regarding the ground-state spin of the quantum dot and the strength of the exchange interaction. This information can be extracted even when mesoscopic fluctuations are taken into account.

The outline of this paper is as follows. In Section \ref{rate-equations} we discuss the rate equation approach for an interacting dot and derive an expression for the sequential thermopower in linear response theory. In particular, we obtain the low-temperature limit of the thermopower for the universal Hamiltonian in terms of the single-particle level widths and excitation energies and spins of a subset of excited states. The low-temperature results are rederived in Section \ref{physical-picture} by calculating the energy transported across the dot per charge carrier. In Section \ref{exchange} we study the fine structure of the sequential thermopower in the presence of exchange correlations and find even-odd effects that are sensitive to the ground-state spin of the dot. In Section \ref{mesoscopic} we demonstrate that these effects survive the presence of mesoscopic fluctuations. In Section \ref{cotunneling} we discuss elastic cotunneling in a dot described by the universal Hamiltonian and its effect on the thermopower. We conclude in Section \ref{conclusion} with a summary and discussion of our main results.

\section{Many-body rate equation calculation of thermopower}\label{rate-equations}

The Beenakker-Staring model for conductance and thermopower is based on a rate equation method in which coherence between the dot and the leads is neglected, and the electron-electron interaction on the dot is represented by a constant charging energy, i.e., the CI model.~\cite{beenakker2}  Hence this approach describes sequential tunneling processes.  The linear response conductance and thermopower are given in terms of the single-particle level transition widths from the left and right leads and the canonical thermal occupation probabilities for the single-particle states on the dot. Later this rate equation approach was generalized by Alhassid et al.~\cite{alhassid1} to describe transitions between arbitrary many-body states of the dot and applied it to the calculation the electrical conductance of a dot that is described by the universal Hamiltonian.~\cite{alhassid3} This Hamiltonian is given by
\begin{equation}\label{univ-H}
\hat{H}=\sum_{\lambda \sigma}\epsilon_{\lambda}a^{\dagger}_{\lambda\sigma}a_{\lambda\sigma}
+\frac{e^{2}}{2C}\hat{N}^{2}-J_s {\bf \hat S}^{2}\;,
\end{equation}
where $a^{\dagger}_\lambda$ creates an electron in energy level $\epsilon_\lambda$ with spin $\sigma$, and
$\hat{N},{\bf \hat S}$ are, respectively,  the total number and total spin operators for the electrons on the dot. The new feature in (\ref{univ-H}) compared to the CI Hamiltonian is the inclusion of the ferromagnetic exchange interaction of strength $J_s$.  Note that for an equally spaced single-particle spectrum, ground states of higher than minimal spin begin to occur when $J_s= 0.5 \,\delta$, whereas full polarization of the dot occurs at $J_s=\, \delta$. Mesoscopic fluctuations cause these thresholds to vary substantially from sample to sample.~\cite{brouwer99, baranger00, alhassid02}  The effect of this exchange term has been studied in detail for the electrical conductance properties,~\cite{alhassid3} but not for the thermopower. More recently, the action for this universal Hamiltonian has been studied in a path integral approach and used to calculate the dot's tunneling density of states and magnetic susceptibilty.~\cite{kiselev, kiselev2}

Here we calculate the thermopower in the Coulomb-blockade regime, in which the charging energy is much larger than the thermal energy, i.e. $e^{2}/C \gg kT$. In this regime, we only need to consider many-body states of the dot with $N$ and $N+1$ electrons, where the value of $N $ is determined by the gate voltage. We consider the case where the thermal energy $kT$ is greater than the average transition width for an electron to tunnel onto the dot. This excludes Kondo-like resonant effects,~\cite{kondo,nguyen09} and allows us to use a rate-equations approach. Following Ref.~\onlinecite{alhassid1}, we consider the rate equations describing the time-evolution of the probability of finding the dot in each many-body states, and look for a steady-state solution. We expand these equations to linear order in the source-drain voltage $V$ and the temperature difference $\Delta T$ across the dot, and arrive at a set of detailed balance equations.

Each detailed balance equation takes the form of a sum over many-particle states; in the case of the pure electrical conductance, (i.e. when $\Delta T=0$), each term in the sum is separately zero,~\cite{alhassid1} provided that orbital occupation number operators commute with the Hamiltonian, as they do for the universal Hamiltonian. This term-by-term solution substantially simplifies the final expression for the electrical conductance. We find that for the thermopower, where $\Delta T \ne 0$, this simplification is no longer exact, unless the transition widths are level-independent. However, when the temperature is much smaller than the mean level spacing (the regime of interest), we find that a term-by-term solution is an excellent approximation, and we use this simplification throughout this work.

We follow the notation of Ref.~\onlinecite{alhassid1}: the $N$ electron states are indexed with $i$ and have energies $\epsilon_i^{(N)}$, while the $N+1$ electron states are indexed with $j$ and have energies $\epsilon_j^{(N+1)}$.  We order them so that the ground states have index $0$. The equilibrium probability of finding the dot in state $i$ is $\tilde{P}_i^{(N)}$, and similarly the probability of finding the dot in state $j$ is $\tilde{P}_j^{(N+1)}$, given by the grand-canonical statistics for the combined manifolds of $N$ and $N+1$ electrons. The transition width from the $N$-electron state $i$ to the $N+1$-electron state $j$ by an electron tunneling from the left (right) lead is denoted by $\Gamma_{ij}^{l(r)}$. The effective Fermi energy in the leads, which includes the effect of the gate voltage, is $\tilde{\epsilon}_{F}$, and the fractional voltage drop across the left junction is $\eta$. We also define  $\epsilon_{ij}=\epsilon_{j}^{(N+1)}-\epsilon_{i}^{(N)} -\tilde{\epsilon}_{F}$ and $f_{ij}= f(\epsilon_{ij})$ where $f(x)=(1+e^{\beta x})^{-1}$ is the Fermi-Dirac function with $\beta=1/kT$. Expanding the occupancy probabilities in linear response theory as $P_{i}^{(N)}=\tilde{P}_{i}^{(N)}\left(1+\Psi_{i}^{(N)}\beta e V+ \Phi_{i}^{(N)} \Delta T/T\right)$ and $P_{j}^{(N+1)}=\tilde{P}_{j}^{(N+1)}\left(1+\Psi_{j}^{(N+1)}\beta e V+ \Phi_{j}^{(N+1)} \Delta T/T\right)$, we find for the thermopower:
\begin{equation}
{\cal S}=\frac{k}{e}\frac{\sum_{ij} \tilde{P}_{i}^{(N)}f_{ij}[ \beta \epsilon_{ij} -(\Phi_{j}^{(N+1)}-\Phi_{i}^{(N)})]\Gamma_{ij}^{l}}{\sum_{ij} \tilde{P}_{i}^{(N)}f_{ij}[ \eta+\Psi_{j}^{(N+1)}-\Psi_{i}^{(N)}]\Gamma_{ij}^{l}} \;.
\end{equation}
There is a set of detailed balance equations for the $\Psi$'s and another set for the $\Phi$'s (see Appendix A). As shown in Ref.~\onlinecite{alhassid1}, the equations for $\Psi$ are satisfied term-by-term for a dot that is described by the universal Hamiltonian. This is not the case for the equations for $\Phi$, but we find that this is an excellent approximation for $T \ll \delta$.  In this approximation, the expression for the thermopower simplifies to~\footnote{For a dot with left-right symmetry (i.e., $\Gamma_{ij}^l = \Gamma_{ij}^r$ for all $i,j$), Eq.~(\ref{thermopowereq}) is exact for a general interaction in the dot (see Appendix A).}
\begin{equation}
{\cal S} \approx -\frac{1}{eT} \frac{\sum_{ij}\tilde{P}_{i}^{N}f_{ij}
\frac{\Gamma_{ij}^{l}\Gamma_{ij}^{r}}{\Gamma_{ij}^{l}+\Gamma_{ij}^{r}}
\epsilon_{ij}}{\sum_{ij}\tilde{P}_{i}^{N}f_{ij}
\frac{\Gamma_{ij}^{l}\Gamma_{ij}^{r}}{\Gamma_{ij}^{l}+\Gamma_{ij}^{r}}}\;.
\label{thermopowereq}
\end{equation}

Using identities for the Fermi-Dirac function, and the relation $\tilde{P}_{i}^{N}=e^{\beta \epsilon_{ij}}\tilde{P}_{j}^{N+1}$, Eq.~(\ref{thermopowereq}) can also be rewritten in the useful alternative form:
\begin{equation}
{\cal S} \approx -\frac{1}{eT} \frac{\sum_{ij}\tilde{P}_{j}^{N+1}(1-f_{ij})
\frac{\Gamma_{ij}^{l}\Gamma_{ij}^{r}}{\Gamma_{ij}^{l}+\Gamma_{ij}^{r}}
\epsilon_{ij}}{\sum_{ij}\tilde{P}_{j}^{N+1}(1-f_{ij})
\frac{\Gamma_{ij}^{l}\Gamma_{ij}^{r}}{\Gamma_{ij}^{l}+\Gamma_{ij}^{r}}} \;.
\label{thermopowereq1}
\end{equation}

We next discuss the low-temperature limit. Both forms (\ref{thermopowereq}) and (\ref{thermopowereq1}) are useful, depending on the sign of $\epsilon_{00}$, i.e., whether the current is particle-like or hole-like. For $\epsilon_{00}>0$, the dot is most likely to have $N$ electrons, and the manifold of $N+1$ electron states is closer in energy than the manifold of $N-1$ electron states. Thus current will flow by transiently adding an electron to the dot, resulting in particle current. In this case, Eq.~(\ref{thermopowereq1}) is more useful. At low temperatures, the largest occupation probability $\tilde P^{(N+1)}_j$ of the $N+1$ electron dot corresponds to the ground state $j=0$ so the dominating terms in the sums of Eq.~(\ref{thermopowereq1}) are the $j=0$ terms. The Fermi-Dirac function approaches a step function at low temperatures, and the sum over initial states $i$ is restricted to the finite number of states with $\epsilon_{i0}>0$, for which  $1-f_{i0}\approx 1$ (this includes the ground state of the $N$ electron dot since we assume $\epsilon_{00}>0$). Thus, at low temperatures and $\epsilon_{00}>0$, the numerator of Eq.~(\ref{thermopowereq1}) is approximated by
\begin{equation}\label{numerator}
\tilde{P}_{0}^{N+1}\sum_{i: \epsilon_{i0}>0}\epsilon_{i0}\frac{\Gamma_{i0}^{l}\Gamma_{i0}^{r}}{\Gamma_{i0}^{l}+\Gamma_{i0}^{r}}\;.
\end{equation}
The occupation probability $\tilde{P}_{0}^{(N+1)}$ can be rather small away from the degeneracy point. However, it cancels out in the thermopower when expression (\ref{numerator}) is combined with a similar expression for the denominator of Eq.~(\ref{thermopowereq1}) to give
\begin{equation}\label{particle-thermopower}
{\cal S} \approx - \frac{1}{eT} \frac{\sum_{i: \epsilon_{i0}>0}\epsilon_{i0}\frac{\Gamma_{i0}^{l}\Gamma_{i0}^{r}}{\Gamma_{i0}^{l}+\Gamma_{i0}^{r}}}{\sum_{i: \epsilon_{i0}>0}\frac{\Gamma_{i0}^{l}\Gamma_{i0}^{r}}
{\Gamma_{i0}^{l}+\Gamma_{i0}^{r}}} \;.
\end{equation}

For $\epsilon_{00}<0$, the dot has $N+1$-electrons, and the manifold of $N$-electron states is closer in energy than the manifold of $N+2$ electron states, so the current will be hole-like. Now we use Eq.~(\ref{thermopowereq}) for the thermopower to find
\begin{equation}\label{hole-thermopower}
{\cal S} \approx -\frac{1}{eT} \frac{\sum_{j: \epsilon_{0j}<0}\epsilon_{0j}\frac{\Gamma_{0j}^{l}\Gamma_{0j}^{r}}{\Gamma_{0j}^{l}+\Gamma_{0j}^{r}}}{\sum_{j: \epsilon_{0j}<0}\frac{\Gamma_{0j}^{l}\Gamma_{0j}^{r}}
{\Gamma_{0j}^{l}+\Gamma_{0j}^{r}}}\;.
\end{equation}

The sequential thermopower in each of expressions (\ref{particle-thermopower}) and (\ref{hole-thermopower}) can be relatively large, even though the electrical and thermal conductances are both small away from degeneracy point.~\footnote{When the competing cotunneling process is included this is no longer the case, and the smallness of $\tilde{P}_{0}^{(N+1)}$ (or $\tilde{P}_{0}^{(N)}$) does reduce the thermopower (see Sec.~\ref{cotunneling}).} This thermopower depends on the gate voltage through the effective Fermi energy $\tilde \epsilon_F$ (which appears in $\epsilon_{ij}$). To see more clearly the dependence on $\tilde \epsilon_F$, we define the excitation energy of the many-particle state $i$ by $\epsilon^{(N)}_{{\rm ex},i}\equiv \epsilon_i^{(N)} - \epsilon^{(N)}_0$ and rewrite $\epsilon_{i0}=\epsilon_{00} - \epsilon_{{\rm ex},i}^{(N)}$. Eq.~(\ref{particle-thermopower}) can then be written as
\begin{equation}\label{particle-thermopower-ex}
{\cal S} \approx -\frac{\epsilon_{00}}{e T} + \frac{1}{eT} \frac{\sum_{i: \epsilon^{(N)}_{{\rm ex},i}< \epsilon_{00}}\epsilon^{(N)}_{{\rm ex},i}\frac{\Gamma_{i0}^{l}\Gamma_{i0}^{r}}{\Gamma_{i0}^{l}+\Gamma_{i0}^{r}}}{\sum_{i: \epsilon^{(N)}_{{\rm ex},i}< \epsilon_{00}}\frac{\Gamma_{i0}^{l}\Gamma_{i0}^{r}}
{\Gamma_{i0}^{l}+\Gamma_{i0}^{r}}} \;.
\end{equation}
Similarly, Eq.~(\ref{hole-thermopower}) can be written as
\begin{equation}\label{hole-thermopower-ex}
{\cal S} \approx -\frac{\epsilon_{00}}{e T} -\frac{1}{eT} \frac{\sum_{j: \epsilon^{(N+1)}_{{\rm ex},j}< |\epsilon_{00}|}\epsilon^{(N+1)}_{{\rm ex},j}\frac{\Gamma_{0j}^{l}\Gamma_{0j}^{r}}
{\Gamma_{0j}^{l}+\Gamma_{0j}^{r}}}{\sum_{j: \epsilon^{(N+1)}_{{\rm ex},j}< |\epsilon_{00}|}\frac{\Gamma_{0j}^{l}\Gamma_{0j}^{r}}
{\Gamma_{0j}^{l}+\Gamma_{0j}^{r}}}\;,
\end{equation}
where we have defined $\epsilon^{(N+1)}_{{\rm ex},j}= \epsilon_j^{(N+1)} - \epsilon^{(N+1)}_0$ and used $\epsilon_{0j}=\epsilon_{00} + \epsilon^{(N+1)}_{{\rm ex},j}$.   Since $\epsilon_{00}= \epsilon_{0}^{(N+1)}-\epsilon_{0}^{(N)} -\tilde{\epsilon}_{F}$, we see that ${\cal S}$ is a piecewise linear function of $\tilde \epsilon_F$ with a slope of $1/e T$ (twice the average classical slope at temperatures $kT \gg \delta$). However, as the effective Fermi energy varies, the number of terms contributing to the sums in Eq.~(\ref{particle-thermopower-ex}) or  Eq.~(\ref{hole-thermopower-ex}) changes, and the thermopower exhibits a discontinuity or a jump. These jumps are in one-to-one correspondence with a subset of the many-body excited states. For gate voltages with $\epsilon_{00}>0$, the jumps in the thermopower correspond to the excited states of the $N$-electron dot that have a non-zero tunneling matrix element with the $N+1$-electron ground state. The maximal allowed excitation is $\epsilon_{00}$ (as dictated by the condition $\epsilon_{i0}>0$). For gate voltages with $\epsilon_{00}<0$, the jumps in the thermopower correspond to the $N+1$-electron excited states that have a non-zero hole tunneling matrix element with the $N$-electron ground state. Such excitations are bounded from above by $|\epsilon_{00}|$ (to satisfy the condition $\epsilon_{0j}<0$).  The distance of the jump from the degeneracy point $\epsilon_{00}=0$ is just the excitation energy of the new state that appears in the sums in either (\ref{particle-thermopower-ex}) or (\ref{hole-thermopower-ex}). As $|\epsilon_{00}|$ increases away from the degeneracy point, the number of excited states contributing to these sums increases in a stepwise manner. The maximal allowed value of $|\epsilon_{00}|$ is $\sim e^2/2C$, hence the largest excitation contributing to the sequential tunneling thermopower is half the charging energy.

We focus our attention on the universal Hamiltonian, for which the orbital occupation numbers and total spin are good quantum numbers. In this case, the excitations that contribute to Eq.~(\ref{particle-thermopower}) move an electron to a level just above the Fermi energy and create a hole below the Fermi energy for an electron to tunnel into. Similarly,  excitations that contribute to Eq.~(\ref{hole-thermopower})  move an electron from the Fermi energy to a level above the Fermi energy.  The excitation spectrum we observe in the sequential thermopower is thus that of the single-particle energy levels (which arise from the non-interacting portion of the universal Hamiltonian) plus any additional spin splitting due to exchange, which differentiates between states with the same orbital occupation numbers.

For a Hamiltonian that is spin-rotation invariant, such as the universal Hamiltonian, the many body states are characterized by good spin quantum numbers $i=(\alpha S M)$, where $S, M$ are the total spin and spin projection quantum numbers, and $\alpha$ denotes all other quantum numbers. The sums over states $i$ and $j$ in Eqs.~(\ref{particle-thermopower}) and (\ref{hole-thermopower}) include a summation over magnetic quantum numbers (including the magnetic quantum number of the ground state $0$) that can be carried out explicitly using the Wigner-Eckart theorem to factorize out the dependence on the magnetic quantum numbers as a Clebsch-Gordan coefficient. For the universal Hamiltonian, orbital occupations are also good quantum numbers and non-zero tunneling matrix elements correspond to many-body states of $N$ and $N+1$ electrons that differ by the occupation of one single-particle orbital $\lambda$. The tunneling width between the $N$-electron state  $i=(\alpha S M)$ and the $N+1$ electron state $j=(\alpha' S' M')$ is then given by
\begin{equation} \label{widths}
\Gamma^{l,r}_{ij}=\frac{1}{2S'+1} (S\; M\; 1/2\; m | S'\; M')^{2}(\alpha' S' || a^\dagger_\lambda || \alpha S )^2 \Gamma^{l,r}_{\lambda}\;
\end{equation}
where $m=M'-M$ is the magnetic quantum number of the electron that tunnels into the dot, and $\Gamma^{l,r}_{\lambda}$ are the widths of level $\lambda$ to decay to the left or right leads. The reduced matrix element $(\alpha' S' || a^\dagger_\lambda || \alpha S )$ (which is independent of the magnetic quantum numbers) is given by Eq.~(4) of Ref.~\onlinecite{alhassid3}. Using Eq.~(\ref{widths}) and the unitarity of the Clebsch-Gordan coefficients, the sums over $M,m,M'$ in Eqs.~(\ref{thermopowereq}) and (\ref{thermopowereq1}) can be carried out explicitly. In particular, the particle-like thermopower of Eq.~(\ref{particle-thermopower})  is now given by

\begin{equation}\label{particle-thermopower-UH}
{\cal S}\approx - \frac{1}{eT} \frac{\sum_{i: \epsilon_{i0}>0}\epsilon_{i0}(2S_i+1)
\frac{\Gamma_{\lambda_i}^{l}\Gamma_{\lambda_i}^{r}}{\Gamma_{\lambda_i}^{l}+
\Gamma_{\lambda_i}^{r}}}{\sum_{i: \epsilon_{i0}>0}(2S_i+1)\frac{\Gamma_{\lambda_i}^{l}\Gamma_{\lambda_i}^{r}}
{\Gamma_{\lambda_i}^{l}+\Gamma_{\lambda_i}^{r}}} \;,
\end{equation}
where $S_i$ is the spin of the intermediate excited state $i$ of the $N$-electron dot characterized by an empty level  $\lambda_i$ below the Fermi energy.
Similarly, the hole-like thermopower of Eq.~(\ref{hole-thermopower}) is given by
\begin{equation}\label{hole-thermopower-UH}
{\cal S}\approx -\frac{1}{eT} \frac{\sum_{j: \epsilon_{0j}<0}\epsilon_{0j}(2S_j+1)
\frac{\Gamma_{\lambda_j}^{l}\Gamma_{\lambda_j}^{r}}{\Gamma_{\lambda_j}^{l}+
\Gamma_{\lambda_j}^{r}}}{\sum_{j: \epsilon_{0j}<0}(2S_j+1)\frac{\Gamma_{\lambda_j}^{l}
\Gamma_{\lambda_j}^{r}}{\Gamma_{\lambda_j}^{l}+ \Gamma_{\lambda_j}^{r}}}\;,
\end{equation}
where $S_j$ is the spin of the intermediate excited state $j$ of the $N+1$ electron dot with a single electron occupying level $\lambda_j$ above the Fermi energy.

\section{Physical origin of the quantum structure in the thermopower}\label{physical-picture}

Our results for the thermopower agree with a simple physical picture we have adapted from Ref.~\onlinecite{staring}, which treats the thermopower in the limit of a quasi-continuous spectrum in the dot. We first use an Onsager relation connecting
the thermopower to the Peltier coefficient $\Pi$:
\begin{equation}
{\cal S}=\frac{\Pi}{T}=\frac{1}{eT}\frac{\partial I_Q}{\partial I} \;.
\end{equation}
Here $\Pi$ is defined as the derivative of the thermal current $I_Q$ with respect to the particle current $I$ under the condition of zero temperature difference $\Delta T=0$. Thus the thermopower can be determined from the heat carried by an electron as it is transported across the dot in a steady-state solution.

We discuss the case $\epsilon_{00}>0$, where the current is particle-like. $\epsilon_{00}$ is the minimal energy required to add an electron to the dot.  However, this energy can arise by any combination of thermal excitation of the electron in the leads and thermal excitation of the dot.  Let the excitation energy of the electron in the left lead be $\Delta_{1}$ and the excitation energy in the dot be $\Delta_2$. These excitations will occur with probabilities proportional to $e^{-\beta \Delta_{1}}$ and $e^{-\beta \Delta_{2}}$, respectively. Since the two excitations occur independently, the probability of both occurring is proportional to $e^{-\beta (\Delta_{1}+\Delta_{2})}$. While $\Delta_{1}$ is, in principle, unbounded, the probability that a total thermal fluctuation is greater than $\epsilon_{00}$ is negligible. Thus, by energy conservation, we must have $\Delta_{1}+\Delta_{2}=\epsilon_{00}$ and these different modes of transport are equally probable. However, only $\Delta_1$ contributes to the Peltier coefficient; this is the energy ultimately transported across the dot when this electron hops off to the right lead.

Since $\Delta_2$ is a dot's excitation energy, and hence quantized, the allowed values of $\Delta_1=\epsilon_{00}-\Delta_{2}$ are also quantized and vary with
$\epsilon_{00}$. Suppose that the excitation energy $\Delta_{2}$ corresponds to exciting the dot to a state $i$, i.e.,  $\Delta_{2}=\epsilon_{i}^{(N)}-\epsilon_{0}^{(N)}$. The heat carried by the electron across the dot is then $\Delta_{1}=\epsilon_{i0}$. To evaluate the Peltier Coefficient, we average over all states $i$ that have an excitation energy less than $\epsilon_{00}$, i.e., all states $i$ with $\epsilon_{i0}>0$. The thermopower is then given by
\begin{equation}
{\cal S} =-\frac{1}{eT}\frac{\sum_{i: \epsilon_{i0}>0}\epsilon_{i0}}{\sum_{i: \epsilon_{i0}>0}1} \;.
\end{equation}
This is exactly the expression found in the rate equation approach, under the assumption that all the transition rates are equal. If we now include a level-dependent weighting factor $\alpha_i$ in the average, to allow for different tunneling rates into the various excited states, we find
\begin{equation}\label{thermopower}
{\cal S} =-\frac{1}{eT}\frac{\sum_{i: \epsilon_{i0}>0}\epsilon_{i0} \alpha_{i}}{\sum_{i: \epsilon_{i0}>0}\alpha_{i}} \;.
\end{equation}
This gives the correct expression (\ref{particle-thermopower}) for the thermopower if we identify $\alpha_{i}=\frac{\Gamma_{i0}^{l}\Gamma_{i0}^{r}}{\Gamma_{i0}^{l}+\Gamma_{i0}^{r}}$, the simplest combination of the partial widths $\Gamma_{i0}^{l}$ and $\Gamma_{i0}^{r}$ that is both symmetric in left and right leads, and vanishes if any of the partial widths is zero.

In the limit of a continuous energy spectrum and uniform transition widths, the values of $\epsilon_{i0}$ are uniformly distributed in the range $[0,\epsilon_{00}]$ and have an average of $\epsilon_{00}/2$. We then find
${\cal S} =-\epsilon_{00}/2eT$,
in agreement with the Beenakker-Staring result in Ref.~\onlinecite{beenakker1} for the thermopower in the ``classical'' limit.~\footnote{We have assumed here  energy-independent transition widths. For realistic dots, one has to average over the mesoscopic fluctuations to recover this limit.}
The case $\epsilon_{00}<0$ can be similarly treated by considering the energy required to remove an electron from the dot, i.e., the energy carried by a hole that tunnels onto the dot.

At low temperatures $kT << \delta$, the energies $\epsilon_{i0}$ (for a given value of $\tilde\epsilon_F$) cannot be treated as uniformly distributed over the allowed range but assume values determined by the respective excitations in the dot. The condition $ \epsilon_{i0}>0$ in the sum of Eq.~(\ref{thermopower}) is equivalent to $\epsilon^{(N)}_{{\rm ex},i} < \epsilon_{00}$. Therefore as the gate voltage (or equivalently the effective Fermi energy $\tilde \epsilon_F$) is varied away from the degeneracy point, $\epsilon_{00}$ increases and the number of terms in the sums of Eq.~(\ref{thermopower}) increases by one each time another excited state is enclosed in the interval $[0,\epsilon_{00}]$. Each of these terms represents an intermediate excited state in the process of moving an electron across the dot. When a particular excited state $i$ becomes energetically allowed (i.e., $\epsilon^{(N)}_{{\rm ex},i} < \epsilon_{00}$) as we go further away from the degeneracy point, we observe a jump in the thermopower.

\section{Exchange and number parity effects}\label{exchange}

The effect of the exchange interaction is to split degenerate spin states that have the same orbital occupation numbers. We will show that this leads to a certain structure of the quantum jumps in the thermopower that depends on the number parity of electrons in the dot. This effect also depends on the ground-state spin of the dot.  We will assume that the ground state of the odd-electron dot is $S=1/2$ and discuss separately the cases where the ground-state spin of the even-electron dot is $S=0$ or $S=1$. We note that the occurrence of an $S=3/2$ ground state (for an odd number of electrons) is much less likely than the occurrence of an $S=1$ ground state (for an even number of electrons) when $J_s <0.5 \,\delta$.

\subsection{Singlet ground state}\label{singlet}

Consider an even-electron dot with a sufficiently small exchange coupling constant so that its ground state has spin $S=0$.  When an excited state of the even-electron dot is created as an intermediate state in the tunneling process discussed in Sec.~\ref{rate-equations}, this state can be either a singlet ($S=0$) or a triplet ($S=1$), depending on the combined spin state of the two singly occupied orbitals.  Since the singlet and triplet states have the same orbital occupations, they are split by an amount $2J_s$, independent of the specific single-particle spectrum. Therefore each of these states (which are degenerate in the absence of exchange) will appear as a quantum jump in the thermopower at values of the gate voltage that are separated by $2J_s$.  In contrast, for an odd-electron dot with a ground state of spin 1/2, the allowed intermediate state will also be of spin 1/2 and no splitting will occur (assuming the even-electron dot has an $S=0$ ground state).  Thus the density of the jumps on the side of even number of electrons is twice as high as the density on the side of odd number of electrons, with pairs of even jumps separated by $2J_s$. This number-parity effect is demonstrated in Fig.~2 for the thermopower of a dot with an exchange interaction of $J_s=0.3 \,\delta$ and for the case of equally-spaced single-particle levels and equal level widths.

The exchange-splitting discussed above is determined by the number parity of electrons in the dot, irrespective of whether the process is particle-like or hole-like. The example shown in Fig.~2 describes a particle-like process on the even side and a hole-like process on the odd side, but similar $2J_s$ splitting occurs for an even-electron dot and a hole-like tunneling process, and there is no splitting for an odd-electron dot and a particle-like process.

\begin{figure}
\includegraphics{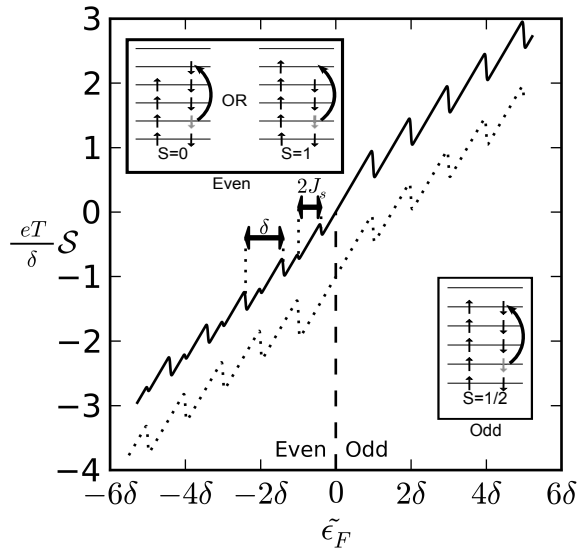}
\caption{Thermopower for a quantum dot with equal level spacings and equal level widths in the presence of exchange interaction $J_s=0.3\,\delta$. Results are shown for $e^{2}/2C=5\delta$ at $kT=\delta/100$. When the number of electrons in the dot is even, intermediate excited states can have either spin $0$ or spin $1$ (see inset in upper left corner; we use the heuristic of aligned spins for spin 1 and antialigned spins for spin 0), leading to an energy splitting of $2 J_s$ for each pair of jumps. The amplitudes of the singlet and triplet jumps are different because of the spin weighting factors in Eq.~(\ref{particle-thermopower-UH}). When the number of electrons in the dot is odd, the intermediate excited states can only have spin $1/2$, and there is no splitting. The dotted line shows the thermopower in the absence of exchange (i.e., $J_s=0$), and has been offset vertically by $eT/\delta$ for clarity. For $\tilde\epsilon_F <0$, with an even number of electrons the process is particle-like and the upper left inset shows excited states of the dot before the tunneling of an electron. For $\tilde\epsilon_F >0$, the process is hole-like; however, for comparison with the even-electron case,  the lower right inset shows an excited state for particle-like transport for an odd-electron dot, in which the excited states have the same spin as the ground state. }
\end{figure}

\subsection{Triplet ground state}\label{triplet}

The lowest $S=0$ state for an even-electron dot is described by double occupancy of its lowest $N/2$ levels. For a sufficiently small exchange coupling constant, this $S=0$ state will be the ground state of the dot.  However, when the energy difference between the $N/2+1$ and $N/2$ orbitals is less than $2J_s$, the dot will have an $S=1$ ground state. In this case, the intermediate excited states in the even-electron dot can still have either $S=0$ or $S=1$, leading to the $2J_s$ splitting in the quantum structure of the thermopower as discussed in Sec.~\ref{singlet}. However, the intermediate excited states of the odd-electron dot have 3 unpaired electrons and can have either $S=1/2$ or $S=3/2$ spin states, and lead to a splitting of $3J_s$ in the quantum structure of the thermopower.
In Fig.~3  we show the appearance of this $3J_s$ splitting for a dot with an equally-spaced spectrum and an exchange constant of $J_s =0.6\, \delta$ (for which the even-electron dot has an $S=1$ ground state). The first jump is not split, because it corresponds to an excitation of an electron from the highest doubly occupied level to the singly  occupied level just above, and the number of unpaired electrons does not change. All other jumps are split by $3J_s$. Thus, when the even-dot ground state is a triplet, the density of jumps is equal on the odd and even sides, and these jumps are paired on both sides. However, the splitting between paired jumps is $2J_s$ on the even side, and $3J_s$ on the odd side.

For an equally-spaced spectrum, the even-electron $S=1$ ground state occurs first (as we increase exchange constant) for  $J_s=0.5\, \delta$, and above that value there is the ``Stoner staircase" of higher spin ground states, leading to even larger values of the exchange splitting.  Typical values of the exchange constant in semi-conductor quantum dots are usually below $J_s= 0.5\,\delta$. However, in the presence of mesoscopic level-spacing fluctuations, there is a finite probability to have
higher-spin ground states at smaller values of $J_s$, particularly $S=1$ for an even-electron dot.  The presence of $3J_s$ splitting is a clear experimental signature of these triplet ground states. We note that the ground-state spin of a dot can in principle be determined by applying an in-plane magnetic field but this is a difficult experiment~\cite{parallel1, parallel2, parallel3} and level crossing at low magnetic field can lead to misidentification of the spin.~\cite{huertas}

\begin{figure}
\includegraphics{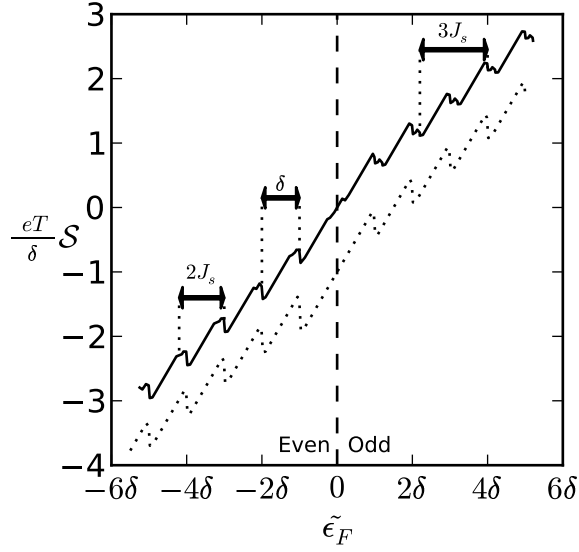}
\caption{Thermopower for a quantum dot with equal level spacings and equal level widths, in the presence of exchange $J_s=0.6\,\delta$. All other parameters are as in Fig.~3. The $2J_s$ splitting for an even number of electrons is still present. Now, however, the ground-state spin of the even-electron dot is $S=1$, causing a $3J_s$ splitting to appear for a dot with an odd number of electrons.  The dotted line shows the thermopower for $J_s=0$, and has been offset vertically by $eT/\delta$ for clarity.}
\end{figure}

\section{Mesoscopic fluctuations}\label{mesoscopic}

The exchange-split quantum jumps are most easily observed at sufficiently low temperatures and when the dot's single-particle levels are equally spaced and have equal tunneling widths. However, in large quantum dots there are mesoscopic fluctuations in the level spacings and widths, and it becomes difficult to identify which jumps are paired together by exchange splitting, and which happen to be close to each other because of the mesoscopic fluctuations in the level spacings. Nevertheless, it is important to note that the splitting persist in the presence mesoscopic fluctuations and, given the particular spin value of the ground state, their value is independent of the particular sample. The jumps will be split by $2J_s$ on the even side of the dot, and, if the ground-state
spin of the even-electron dot is $S=1$, by $3J_s$ on the odd side, regardless of level fluctuations (ignoring samples for which the ground-state spin is larger than $S=1$).  Fluctuations in the transition width for each level will affect the size of the jumps, but will not change their position, and thus the jumps will still be separated by $2J_s$ or $3J_s$. In Fig.~4 we show the thermopower for a dot with the same RMT set of single-particle levels and widths as in Fig.~1b, but now including an exchange interaction with a strength of $J=0.3\,\delta$. The greater density of jumps on the even side is apparent, but without prior knowledge of the energy levels, it is difficult to identify which jumps on the even side are paired.   Hence a statistical analysis is needed. Such analysis becomes substantially more complex due to the effects of cotunneling (see Sec.~\ref{cotunneling}), and appears to be difficult to carry out with current experimental techniques.

\begin{figure}
\includegraphics{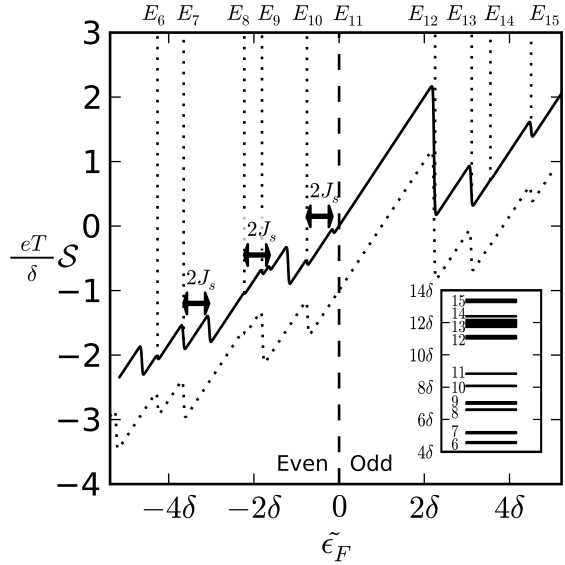}
\caption{Thermopower for a quantum dot whose single-particle energy levels and level widths are sampled from RMT, and for an exchange constant of $J_s = 0.3 \,\delta$. Results are shown for $e^{2}/2C=5\delta$ at $kT= \delta/100$. The energy levels have been labeled, and the inset shows the spectrum with the width of each line proportional to the conductance through that level. The dotted curve, offset for clarity, is the thermopower for the same RMT sample but without an exchange interaction ($J_s=0$). For this particular sample, the ground-state spin of the even-electron dot is a singlet, so we observe $2J_s$ splitting on the even side and no splitting on the odd side. Note that the width fluctuations have removed a jump corresponding to $E_{14}$.}
\end{figure}

An interesting statistical quantity is the distribution of the spacing between neighboring jumps in the thermopower.  These spacings will be smoothly distributed when they arise from different single-particle levels (because of level spacing fluctuations). However, jump spacings that arise from exchange-split jumps, are expected to lead to a large spike at $\sim 2J_s$ and a smaller spike at $\sim 3J_s$. Here we assume $J_s < 0.5\,\delta$, so spin $S=1$ ground states are not too frequent and higher spin ground states are rare. A histogram of this distribution is shown in Fig.~5. It is constructed from the thermopower line shapes of different samples drawn from the gaussian orthogonal ensemble and confirm our expectations. Presently, it is unrealistic to observe such a distribution experimentally, since in the presence of cotunneling, it is difficult to observe more than one quantum jump (see Sec.~\ref{cotunneling}). However, if the measurement of small voltages across a quantum dot can be substantially improved to allow for a more isolated dot (through which the conductance is smaller), then it might be possible to measure several such jumps before the contunneling cutoff sets in. Due to the logarithmic dependence of this cotunneling cutoff on the conductance, one would have to measure conductances that are several orders of magnitude smaller in order to push the cotunneling cutoff several level spacings away from the degeneracy point. Recent advances in using capacitively coupled quantum point contacts to measure extremely small currents~\cite{ensslin08} suggest that perhaps such an experiment might be feasible in the future.

\begin{figure}
\includegraphics{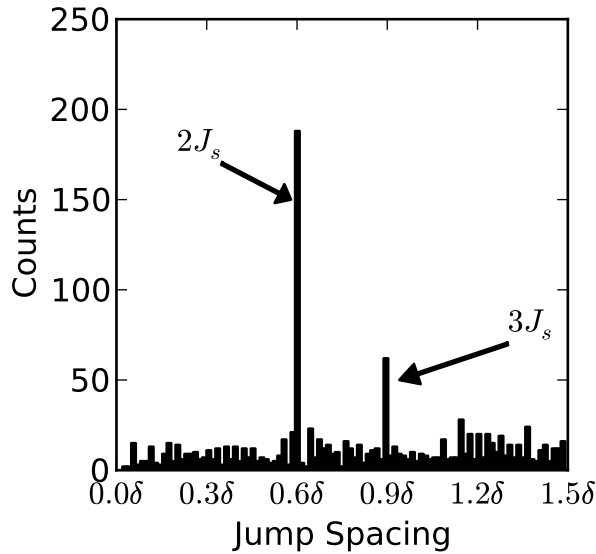}
\caption{Histogram of jump spacings generated by sampling single-particle levels and level widths from the gaussian orthogonal
ensemble, and evaluating the corresponding thermopower line shapes from Eq.~(\ref{thermopowereq}) for an exchange constant of $J_s=0.3\,\delta$. This distribution has two peaks at $2J_s$ and $3 J_s$, whose relative heights provides a measure of the probability of spin $S=1$ ground states at this value of $J_s$. These peaks in the jump spacing distribution cannot be observed with current experimental techniques (because of cotunneling effects), but this might change if ultra-low voltage measurements become feasible in the future.}
\end{figure}

\section{Cotunneling}\label{cotunneling}

The rate equation approach of Sec.~\ref{rate-equations} assumes that the dominant transport process across the dot is sequential tunneling. However, at low temperatures, and away from the Coulomb-blockade conductance peaks, this is not the case. Cotunneling, the tunneling of electrons across the dot through virtual excitations (rather than thermal excitations), becomes increasingly important as the temperature is reduced. There are two cotunneling processes: inelastic cotunneling describes the virtual tunneling of an electron into and out of the dot that leaves the dot in a different state (with the same number of electrons), and elastic cotunneling that leaves the dot in the same state.

The thermopower is given by the ratio $G_{T}/G$, where $G={\partial I}/{\partial \Delta V}|_{\Delta T=0}$, and $G_T={\partial I}/{\partial \Delta T}|_{V=0}$ are, respectively, the electrical conductance and thermal conductance coefficients. Making the approximation that the various contributions to the conductance and thermal conductance are additive, we have~\cite{koch04}
\begin{equation}
{\cal S} \approx \frac{G_{T}^{\rm sequential}+G_{T}^{\rm elastic}+G_{T}^{\rm inelastic}}
{G^{\rm sequential}+G^{\rm elastic}+G^{\rm inelastic}} \;.
\end{equation}

In the middle of the conductance valley, $G_{T}^{\rm sequential}$ and $G^{\rm sequential}$ are proportional to $\exp(- e^{2}/2C kT)$, and are thus both are very small at low temperatures. If sequential tunneling is the only transport process, the exponential factor cancels between numerator and denominator, and the thermopower is not suppressed in the conductance valley. Including cotunneling increases both $G_{T}$ and $G$. However, the relative increase in $G$ is greater than that in $G_{T}$, and the thermopower is suppressed when cotunneling is significant.

At the low temperatures $kT\ll \delta$ necessary to see the quantum structure in the thermopower, elastic cotunneling processes dominate inelastic cotunneling processes.~\cite{cronenwett} The elastic cotunneling conductance $G^{\rm elastic}$ in the presence of exchange correlations is calculated in Appendix B. Taking the example of an even-electron dot with an $S=0$ ground state, we find the average elastic cotunneling conductance to be
\begin{equation}\label{average-G-co}
\bar{G}^{\rm elastic}=\frac{\hbar G^{l} G^{r} \delta}{2 \pi e^{2}}\left(\frac{1}{\tilde{\epsilon}_{F}+E_c -\frac{3}{4} J_s}-\frac{1}{\tilde{\epsilon}_{F}+ \frac{3}{4} J_s}\right)\;,
\end{equation}
where $G^{l(r)}=e^2 \nu_d \Gamma_0^{l(r)}$ is the conductance through the left (right) tunnel junction (with $\nu_d$ being the single-particle density of states in the dot per unit area), $E_c=e^2/C$ and we measure the effective Fermi energy relative to the degeneracy point.  For $J_s=0$, Eq.~(\ref{average-G-co}) reduces to the known expression for the average elastic cotunneling conductance in the CI model.~\cite{averin,aleiner,ABG02,glazman02} In Appendix B we also calculate the average elastic cotunneling thermal conductance (for a dot with zero ground-state spin) to be
\begin{equation}\label{average-GT-co}
\bar G_{T}^{\rm elastic}=\frac{\pi}{6 e} \frac{\hbar}{e^{2}}G^{l}G^{r} k^{2}T \delta \left[\frac{1}{(\tilde{\epsilon}_{F} + \frac{3}{4}J_s)^{2} }-\frac{1}
{(\tilde{\epsilon}_{F}+E_{c} - \frac{3}{4}J_s)^{2}}\right]\;.
\end{equation}
In the middle of the valley $\tilde\epsilon_F=-E_c/2$, and $\bar G_{T}^{\rm elastic}=0$ as is the case in the absence of exchange. In general, we see from Eqs.~(\ref{average-G-co}) and (\ref{average-GT-co}) that the effect of exchange correlations on elastic cotunneling is small.

As we move away from the degeneracy point, cotunneling dominates the sequential tunneling, and it becomes very difficult to observe the quantum jumps discussed in Sec.~\ref{rate-equations}. Thus, even though the quantum jumps are contained in $G_{T}$, the relevant cutoff for their observation occurs when $\bar G^{\rm sequential} \sim \bar G^{\rm elastic}$. Using $\bar G^{\rm sequential}=  \frac{\delta}{kT} e^{-|\tilde\epsilon_{F}|/kT} \frac{G^l G^r} { G^{l}+G^{r}}$, this condition reads (for $J_s=0$)
\begin{equation}
\frac{|\tilde \epsilon_F|}{\delta} \sim \frac{kT}{\delta} \ln\left[ \frac{2\pi e^{2}/\hbar}{G^{l}+G^{r}} \left({\tilde\epsilon_F \over kT}\right) \right]\;.
\end{equation}

For $G^{l}=G^{r}=10^{-3}e^{2}/\hbar$, which are measurable values in current experiments, and $kT=\delta/15$ (about the largest temperature at which the quantum fine structure can still be observed), we find $\tilde\epsilon_F \sim 0.6\,\delta-0.7\,\delta$. This puts into question the possibility of observing {\it pairs} of jumps split by $2J_s$, the simplest signature of exchange interactions.
First, the cut-off implies that the pairs which could be observed are those nearest the degeneracy point, but, as can be seen from Eq.~(\ref{particle-thermopower}), the {\it amplitude} of the jump closest to the degeneracy point is determined by the lowest excitation energy in the dot and is thus quite small.
Thus samples with smaller $J_s$, which would avoid the cotunneling cutoff, will also show smaller jumps. Moreover such jumps will be rounded at finite $T$ and our simulations indicate that in practice one cannot resolve jumps that lie within a couple of  $k T$ of the degeneracy point.  While there are configurations of levels with which one {\it can} observe pairs of jumps, their occurrence is very rare.
Thus, with current experimental methods, the cotunneling cutoff will make the paired jump signature of the exchange interaction difficult to measure.

While this simplest signature is thus a challenge for future experiments, there is a less direct method for observing the effect of exchange correlations on the thermopower which is quite feasible with current experimental techniques. The net effect of the exchange interaction on the many-body spectrum is to increase the density of low-energy excited states.  This is because higher spin-states which cost additional confinement energy are brought down near the ground-state by ferromagnetic exchange correlations.  Alternatively the exchange interaction can make the ground state a higher spin state, leaving a lower spin excited state very near the ground state (as happens in the regime of the mesoscopic Stoner transition).  The result is that the probability of observing even a {\it single} quantum jump in the thermopower (as has been achieved already experimentally in Ref.~\onlinecite{dzurak2}) is substantially enhanced by ferromagnetic exchange correlations.  Hence it is possible to measure the probability of occurrence of observable quantum jumps in the thermopower and infer from that the value of $J_s$.

To be definite, cotunneling makes it impossible to observe jumps outside a certain cutoff $C$ away from the degeneracy point (with current experimental methods $C \sim 0.6\,\delta$ using $kT=\delta/15 $ and $G \sim 10^{-3}e^{2}/\hbar$). In addition, the smallness and rounding of jumps near the degeneracy point put a lower cutoff $c\sim 0.2\,\delta$ on how close to the degeneracy point a jump can be clearly seen. Thus, a jump can be observed if the actual first excited state is in the interval $\sim (0.2\, \delta, 0.6\,\delta)$. In the absence of exchange correlations, the probability $p$ of such an observable first excited state is $p=\int^{C/\delta}_{c/\delta}p_{W}(s)ds$, where $p_W(s)=(\pi/2) s e^{-\frac{\pi}{4}s^{2}}$  is Wigner's surmise for the nearest-neighbor level spacing distribution in the absence of a magnetic field. Thus, the likelihood of observing a jump in the even-electron dot is
$p=e^{-\frac{\pi}{4}(c/\delta)^2}-e^{-\frac{\pi}{4}(C/\delta)^2} \approx 0.22$.  In Fig.~6(a) we show that exchange correlations enhance this probability $p$ dramatically, by more than a factor of two for typical values of $J_s \approx 0.3$, and as much as a factor of three for $J_s \approx 0.6$,

\begin{figure}
\includegraphics{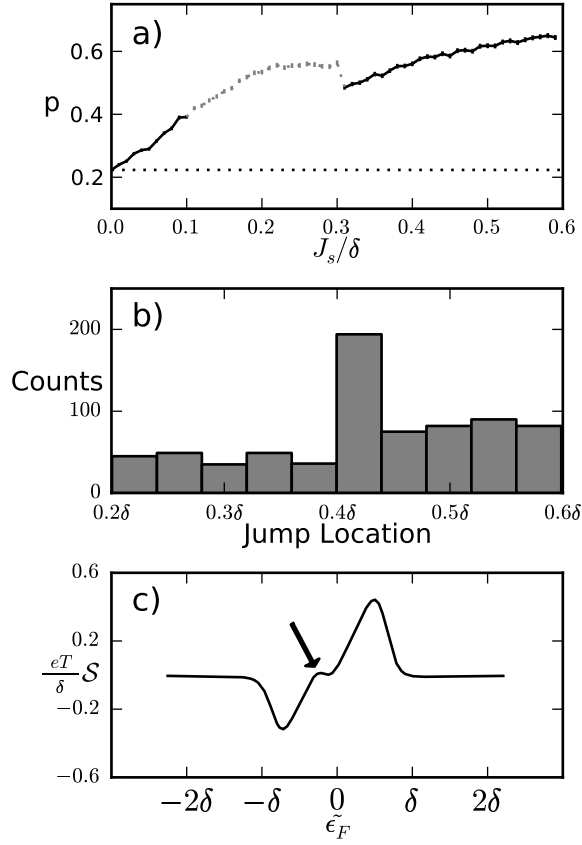}
\caption{(a) The probability $p$ of observing a quantum jump in the thermopower in the interval $(c,C)=(0.2\,\delta,0.6\,\delta)$ versus the exchange coupling constant $J_s/\delta$.  The curve is monotonic once the interval (shown in gray) $(c/2,C/2)=(0.1\,\delta,0.3\,\delta)$ is excluded, and can thus be used to determine $J_s$ experimentally when $J_s$ is outside this interval. The exchange interaction leads to a strong enhancement of $p$ as compared with its value in the absence of exchange (dotted line). (b) A histogram of quantum jumps that are observed within the interval $(c,C)=(0.2\,\delta,0.6\,\delta)$ for an exchange strength of $J_s=0.2\,\delta$. The large spike at $2 J_s$ can be used to determine $J_s$ when $c/2<J_s<C/2$. (c) A thermopower trace (in the presence of elastic cotunneling) for a particular sample in which a triplet quantum jump (indicated by the arrow) is observed.}
\end{figure}

As already noted, the basic origin of this enhancement is the increase in the number of low-energy excited states due to the reduction in energy of higher spin states in the presence exchange correlations.  To calculate this enhanced probability $p$ for $J_s\neq 0$ we identify the lowest excited states of the $N$ electron dot that have an overlap with the $N+1$ electron ground state (after an electron tunnels into the dot), and determine whether any of these excitations fall within the observability interval $(c,C)$. We need to include the possibility of both singlet and triplet ground states, which slightly complicates the analysis. For sufficiently small $J_s$, we can choose a random matrix spectrum and pick two adjacent spacings $s$ and $t$ (whose ensemble average is $\delta$) to represent the first and second single-particle spacings above the Fermi level of the dot. If $s >2J_s$, then the ground state is a singlet $S=0$, and the lowest excited states are the triplet with an excitation energy of $s-2J_s$ and a singlet at excitation of $s$.  In contrast, when $s<2J_s$, the ground state becomes a triplet and there are four relevant excited states with energies $2J_s - s, t, 2J_s, 2J_s + t$ all of which can cause jumps.  However the $2J_s$ excited state, which is just the singlet partner of the triplet ground state (i.e. same occupation numbers, but different spin), will always give a jump at a distance $2J_s$ from the degeneracy point, independent of level fluctuations (as long as $s<2J_s$).  In fact this is a jump pair of the type we described in Sec.~\ref{exchange}, but with the lower energy state corresponding to the ground state, making it unobservable since it carries no energy through the dot.  Hence for $J_s$ in the interval  $c/2 <J_s< C/2$ (i.e., $0.1\,\delta<J_s< 0.3\,\delta$) the quantum jump that corresponds to the $2J_s$ excited state is not suppressed by cotunneling effects, and a simple histogram of the jump location (measured from the degeneracy point) will exhibit a peak at $2J_s$, allowing one to read off the value of the exchange constant. This is shown in the Fig.~6(b). The overall probability of observing a quantum jump in the interval $(c,C)$ for this determined value of $J_s$ is given by the plot in Fig.~6(a) and can be used as a check of the result.

In general, at larger values of $J_s$, ground-state spins higher than $S=1$ have non-negligible probability, and for each random matrix sample we have considered all possible states of an even-electron dot that contribute to the thermopower and whose excitation energy is in the interval $(c,C)$.
The probability $p(J_s)$  (within the range $0 < J_s < 0.6\, \delta$) is given in Fig.~6(a).  The interval $ 0.1 < J_s <0.3$ is shown in gray; in this interval a simple histogram will determine $J_s$ as shown in Fig.~6(b) for $J_s=0.2\,\delta$.  Outside of this interval of $J_s$, dots with ground state spin $S=1$ will not have the $2J_{s}$ jump within the window of observability, and there will be no peak in the histogram, but rather a smooth distribution due to level fluctuations. However, the total probability of observing a jump can be used to infer the value of $J_s$, since $p$ is a {\it monotonic} function of $J_s$ outside of the gray-scale region, and the corresponding value of $J_s$ can be read off from this curve.  This method for determining $J_s$ is experimentally feasible. The required ensemble of thermopower traces can be obtained by using finger gates and a back gate voltage. In Fig.~6(c) we show an example of a thermopower trace that exhibits a single quantum jump. Jumps are more likely to occur on the even side of the dot (this is the case in the example of Fig.~6), a feature that might be useful for determining the number parity of electrons on the dot with reasonable certainty by measuring the thermopower trace over several Coulomb blockade oscillations.

\section{Conclusion}\label{conclusion}

We have shown how the presence of ferromagnetic exchange correlations modifies the fine quantum structure of the thermopower of a many-electron quantum dot that is described by the universal Hamiltonian. The quantum structure in the thermopower is sensitive to the excitation energies and transition widths of a subset of excited many-body states in the dot. In general, the exchange interaction splits the quantum jumps by an integer times the exchange constant, independent of mesoscopic level and width fluctuations.  For the specific case when the ground state has spin $S=0$ for an even number of electrons and $S=1/2$ for an odd number of electrons, there are twice as many jumps in the even valleys compared with the odd valleys, a signature of number parity.  In principle, a histogram of the spacing between neighboring jumps can be used to measure the exchange constant in the system and also determine the probability of a triplet ground state. Cotunneling effects suppress these strong signatures under current experimentally realizable conditions, and the observation of these signatures would require the ability to measure much smaller conductances in almost-isolated dots.  However, exchange correlations increase significantly the probability of observing any quantum jump in the thermopower near the degeneracy point (in the presence of cotunneling), and this is likely the reason that such jumps were observed.  A detailed study of the distribution of the jump energy within a certain observability window and of the probability of occurrence of such jumps, can therefore be used to estimate the exchange constant $J_s$ (using current experimental methods) and determine the number parity of the electrons on the dot.  Note that this determination does not require a magnetic field, eliminating one of the difficulties in measuring spin effects in quantum dots by attempting to apply a purely in-plane field.

\begin{acknowledgements}

We acknowledge L.I. Glazman, M.N. Kiselev and K.A. Matveev for useful discussions. This work was supported in part by the B. Edward Bensinger Prize, by the NSF grant DMR-0908437 and by the U.S. DOE Grant No.~DE-FG-0291-ER-40608.
\end{acknowledgements}

\appendix

\section{Detailed balance equations}

The dot is described by the following rate equations~\cite{alhassid1}
\begin{equation}
\frac{\partial P_{i}^{(N)}}{\partial t}=\sum_{j}P_{j}^{(N+1)}[(1-f_{ij}^{l})\Gamma_{ij}^{l}+(1-f_{ij}^{r})\Gamma_{ij}^{r}]-P_{i}^{(N)}\sum_{j}[f_{ij}^{l}\Gamma_{ij}^{l}+f_{ij}^{r}\Gamma_{ij}^{r}] \qquad \forall i \;,
\label{rate1}
\end{equation}
\begin{equation}
\frac{\partial P_{j}^{(N+1)}}{\partial t}=\sum_{i}P_{i}^{(N)}[f_{ij}^{l}\Gamma_{ij}^{l}+f_{ij}^{r}\Gamma_{ij}^{r}]-P_{j}^{(N+1)}\sum_{i}[(1-f_{ij}^{l})\Gamma_{ij}^{l}+(1-f_{ij}^{r})\Gamma_{ij}^{r}] \qquad \forall j \;.
\label{rate2}
\end{equation}
In the presence of a potential difference $eV$ and a temperature difference $\Delta T$ between the two leads, the Fermi-Dirac functions at the left and right lead are given by
\begin{subequations}
\begin{eqnarray}
f_{ij}^{l}=\left[1+e^{(\epsilon_{ij}+\eta eV)/k(T+\Delta T)}\right]^{-1}\;,\\
\label{fermi1}
f_{ij}^{r}=\left[1+e^{(\epsilon_{ij}-(1-\eta) eV)/kT}\right]^{-1} \;,
\label{fermi2}
\end{eqnarray}
\end{subequations}
where $\eta$ is the fractional voltage drop across the left barrier. We are interested in a steady-state solution, i.e., $\partial P_{i}^{(N)}/ \partial t=\partial P_{j}^{(N+1)} / \partial t=0$ for all $i$ and $j$. In linear response, where $V$ and $\Delta T$ are small, we expand the probabilities $P_i^{(N)}$ and $P_j^{(N+1)}$ to first order in $eV$ and $\Delta T$ around the respective (grand-canonical) equilibrium probabilities $\tilde P_i^{(N)}$ and  $\tilde P_j^{(N+1)}$
\begin{subequations}\label{linear-P}
\begin{eqnarray}
P_{i}^{(N)} & = &\tilde{P}_{i}^{(N)}(1+\Psi_{i}^{(N)}\beta e V+ \Phi_{i}^{(N)} \Delta T/T) \;,\\
P_{j}^{(N+1)}& = &\tilde{P}_{j}^{(N+1)}(1+\Psi_{j}^{(N+1)}\beta e V+ \Phi_{j}^{(N+1)} \Delta T/T) \;,
\end{eqnarray}
\end{subequations}
where $\beta=1/kT$.  Expanding the Fermi-Dirac functions to first order in $eV$ and $\Delta T$, we find
\begin{subequations}\label{linear-f}
\begin{eqnarray}
f_{ij}^{l} & = &  f_{ij}+\eta eV f_{ij}' - \Delta T \frac{\beta \epsilon_{ij}}{T} f_{ij}' \;,\\
f_{ij}^{r} & = & f_{ij}-(1-\eta)eV f_{ij}' \;,
\end{eqnarray}
\end{subequations}
where  $f_{ij}=f(\epsilon_{ij})=[1+e^{\beta \epsilon_{ij}}]^{-1}$, and $f_{ij}'$ denotes differentiation with respect to energy.

Inserting Eqs.~(\ref{linear-P}) and (\ref{linear-f}) into the right-hand sides of Eqs.~(\ref{rate1}) and (\ref{rate2}) (setting the left-hand sides to zero), and applying further simplifications as in Ref.~\onlinecite{alhassid1}, we find two set of linear equations. The set for $\Psi$'s
\begin{subequations}\label{linear-psi}
\begin{eqnarray}
\sum_{j}f_{ij}[(\Gamma_{ij}^{l}+\Gamma_{ij}^{r})(\Psi_{j}^{(N+1)}-\Psi_{i}^{(N)})+ (\eta \Gamma_{ij}^{l}-(1-\eta)\Gamma_{ij}^{r})] = 0 \quad\forall i \;, \\
\sum_{i}(1-f_{ij})[(\Gamma_{ij}^{l}+\Gamma_{ij}^{r})(\Psi_{j}^{(N+1)}-\Psi_{i}^{(N)})+ (\eta \Gamma_{ij}^{l}-(1-\eta)\Gamma_{ij}^{r})] = 0  \quad\forall j
\end{eqnarray}
\end{subequations}
is identical to the set derived in Ref.~\onlinecite{alhassid1}.  The new set of equations for the $\Phi$'s is
\begin{subequations}\label{linear-phi}
\begin{eqnarray}
\sum_{j}f_{ij}[(\Gamma_{ij}^{l}+\Gamma_{ij}^{r})(\Phi_{j}^{(N+1)}-\Phi_{i}^{(N)})-\beta \epsilon_{ij} \Gamma_{ij}^{l}]=0 \quad\forall i \;, \\
\sum_{i}(1-f_{ij})[(\Gamma_{ij}^{l}+\Gamma_{ij}^{r})(\Phi_{j}^{(N+1)}-\Phi_{i}^{(N)})-\beta \epsilon_{ij}\Gamma_{ij}^{l}]=0 \quad\forall j \;.
\end{eqnarray}
\end{subequations}

The electrical current through the left lead is given by
\begin{equation}
I=\frac{e}{\hbar}\sum_{ij}[P_{i}^{(N)}f_{ij}^{l}-P_{j}^{(N+1)}
(1-f_{ij}^{l})]\Gamma_{ij}^{l} \;,
\end{equation}
and in linear response
\begin{equation}\label{left-current}
I=\frac{e}{\hbar}\sum_{ij} \tilde{P}_{i}^{(N)}f_{ij}\left[ \frac{\Delta T}{T} (\beta \epsilon_{ij} -(\Phi_{j}^{(N+1)}-\Phi_{i}^{(N)}))-\beta eV(\eta+(\Psi_{j}^{(N+1)}-\Psi_{i}^{(N)}))\right]\Gamma_{ij}^{l} \;.
\end{equation}
Setting $I=0$, we obtained our main result for the thermopower ${\cal S} =V/\Delta T$
\begin{equation}
{\cal S}=\frac{k}{e}\frac{\sum_{ij} \tilde{P}_{i}^{(N)}f_{ij}[ (\beta \epsilon_{ij} -(\Phi_{j}^{(N+1)}-\Phi_{i}^{(N)})]\Gamma_{ij}^{l}}{\sum_{ij} \tilde{P}_{i}^{(N)}f_{ij}[ \eta+\Psi_{j}^{(N+1)}-\Psi_i^{(N)}]\Gamma_{ij}^{l}} \;.
\end{equation}

If the detailed balance equations (\ref{linear-psi}) and (\ref{linear-phi}) are satisfied term-by-term, the thermopower can be written as in Eq.~(\ref{thermopowereq}). For the universal Hamiltonian, Eqs.~(\ref{linear-psi}) are satisfied term-by-term~\cite{alhassid1} but this does not generally hold for Eqs.~(\ref{linear-phi}).

Another general expression for ${\cal S}$ can be obtained by calculating the current through the right lead.~\cite{koch04} For $\eta=1$
\begin{equation}\label{right-current}
I=\frac{e}{\hbar}\sum_{ij} \tilde{P}_{i}^{(N)}f_{ij}\left[ \frac{\Delta T}{T} (\Phi_{j}^{(N+1)}-\Phi_{i}^{(N)})+\beta eV(\Psi_{j}^{(N+1)}-\Psi_{i}^{(N)})\right]\Gamma_{ij}^{r} \;.
\end{equation}
Since the currents in the left and right leads must be equal, we can take the average of Eqs.~(\ref{left-current}) (at $\eta=1$) and (\ref{right-current}) to find
\begin{equation}\label{left-right-S}
{\cal S}=-\frac{k}{e}\frac{\sum_{ij} \tilde{P}_{i}^{(N)}f_{ij}[\beta \epsilon_{ij} \Gamma_{ij}^{l} - (\Phi_{j}^{(N+1)}-\Phi_{i}^{(N)})(\Gamma_{ij}^{l} - \Gamma_{ij}^{r})]}{\sum_{ij} \tilde{P}_{i}^{(N)}f_{ij}[ \Gamma_{ij}^{l}+ (\Psi_{j}^{(N+1)}-\Psi_i^{(N)})(\Gamma_{ij}^{l}- \Gamma_{ij}^{r})]} \;.
\end{equation}
A simplified expression for the thermopower followed for a dot that has left-right symmetry,~\cite{koch04} i.e., $\Gamma_{ij}^{l}=\Gamma_{ij}^{r}$ for all $i$ and $j$. In this case expression (\ref{left-right-S}) reduces to
\begin{equation}\label{symmetric-S}
{\cal S}=-\frac{1}{eT} \frac{\sum_{ij}\tilde P_i^{(N)}f_{ij} \epsilon_{ij}\;\Gamma_{ij}^{l}}{\sum_{ij}\tilde P_i^{(N)}f_{ij}
\;\Gamma_{ij}^{l}}\;.
\end{equation}
For a symmetric dot $\frac{\Gamma_{ij}^{l} \Gamma_{ij}^{r}}{\Gamma_{ij}^{l} + \Gamma_{ij}^{r}}=\frac{1}{2}\Gamma_{ij}^{l}$, and Eq.~(\ref{symmetric-S}) is equivalent to Eq.~(\ref{thermopowereq}). Thus for a dot with left-right symmetry, Eq.~(\ref{thermopowereq}) holds generally, even when the detailed balance equations are not satisfied term-by-term.

\section{Electrical and thermal conductances for elastic cotunneling}

In this appendix we calculate the elastic cotunneling thermal conductance $G_{T}^{\rm elastic}$. The tunneling Hamiltonian is
\begin{equation}
H_{\rm tun}=\sum_{k,a}t_{ka}\psi^{\dagger}({\bf r}_{a})c_{k,a}+h.c. \;,
\end{equation}
where $\psi^{\dagger}({\bf r}_{a})$ is the dot's field operator creating an electron at the point contact ${\bf r}_{a}$ ($a=r,l$), $t_{ka}$ is the tunneling amplitude for an electron in lead $a$ with momentum $k$, and $c_{k,a}$ are annihilation operators in the leads.

Following Ref.~\onlinecite{koch04}, we denote the cotunneling transition rate from lead $a$ to lead $b$ as the dot makes a transition from state $(N,i)$ to state $(N,i')$ by $W_{ii'}^{ab}$. We have
\begin{equation}
W_{ii'}^{ab}=\int d\epsilon_{a}f^{a}(\epsilon_{a})[1-f^{b}(\epsilon_{b})]\Gamma_{ii'}^{ab} \;,
\end{equation}
where $\epsilon_{b}=\epsilon_{a}+\epsilon_{i}^{(N)}-\epsilon_{i'}^{(N)}$
For elastic cotunneling, $i=i'$ and $\epsilon_{b}=\epsilon_{a}$.

We calculate the elastic transition width $\Gamma^{ab}_{ii}$ in second order perturbation theory
\begin{eqnarray}\label{elastic-Gamma}
\Gamma_{ii}^{ab}=\frac{2 \pi}{\hbar}\sum_{k,k'} \delta(\epsilon_{k}-\epsilon_{a}) \delta(\epsilon_{k'}-\epsilon_{b}) \left| \sum_{j}\frac{t_{k'}^{b*} \langle N, i | \psi({\bf r}_{b})|N+1, j \rangle t_{k}^{a} \langle N+1, j|\psi^{\dagger}({\bf r}_{a})|N, i \rangle}{\epsilon_{j}^{(N+1)}-\epsilon_{i}^{(N)}-\tilde{\epsilon}_{F}-\epsilon_{a}}
\right. \\ \nonumber \left. - \sum_{j}\frac{t_{k}^{a} \langle N, i | \psi^{\dagger}({\bf r}_{a})|N-1, j \rangle t_{k'}^{b*} \langle N-1, j|\psi({\bf r}_{b})|N, i \rangle}{\epsilon_{i}^{(N)}-\epsilon_{j}^{(N-1)}-
\tilde{\epsilon}_{F}- E_c -\epsilon_{b}}\right|^{2} \;,
\end{eqnarray}
where $E_c=e^2/C$. The first contribution to the amplitude corresponds to an electron-like current; the second corresponds to a hole-like current. Note that $\Gamma_{ii}^{rl}=\Gamma_{ii}^{lr}$.

The cotunneling current is~\cite{koch04}
\begin{equation}
I=e \sum_{i}\tilde P_{i}^{(N)}(W_{ii}^{lr}-W_{ii}^{rl}) \;,
\end{equation}
where
\begin{equation}
W_{ii}^{lr}-W_{ii}^{rl}=\int d\epsilon [f^{l}(\epsilon)-f^{r}(\epsilon)]\Gamma_{ii}^{lr} \;.
\end{equation}
Expanding the Fermi-Dirac functions $f^{l,r}$ in $\Delta T$ and $eV$, we find the electrical and thermal eleatic cotunneling conductances to be
\begin{subequations}\label{elastic-G}
\begin{eqnarray}
G^{\rm elastic} & = & e^2 \sum_{i}\tilde P_{i}^{(N)}\int d\epsilon  f'(\epsilon) \Gamma_{ii}^{lr} (\epsilon)\;, \\
G^{\rm elastic}_{T} & = &-e\sum_{i}\tilde P_{i}^{(N)}\int d\epsilon \frac{\epsilon}{T} f'(\epsilon) \Gamma_{ii}^{lr} (\epsilon) \;.
\end{eqnarray}
\end{subequations}
In the limit $kT \ll \delta$, $f'(\epsilon) \to -\delta(\epsilon)$ and Eqs.~(\ref{elastic-G}) reduce to
\begin{subequations}\label{elastic-low-T}
\begin{eqnarray}
G^{\rm elastic} & = & -e^2 \tilde P_i^{(N)} \Gamma_{ii}^{ lr} (0) \;, \\
G_{T}^{\rm elastic}& = & e k^{2} T \frac{\pi^{2}}{3} \tilde P_{i}^{(N)} \partial \Gamma_{ii}^{lr}/\partial \epsilon \left. \right|_{\epsilon=0} \;,
\end{eqnarray}
\end{subequations}
where the expression for $G_{T}^{\rm elastic}$ is obtained after
expanding $\Gamma_{ii}^{lr}(\epsilon)$ to first order in $\epsilon$.
The thermopower obtained from Eqs.~(\ref{elastic-low-T}) is in agreement with Mott's rule although the latter is derived for a non-interacting electron gas (see, e.g., in Ref.~\onlinecite{ziman72}). Our results hold in the presence of interactions in the dot.

We focus on the universal Hamiltonian for which the matrix elements in Eq.~(\ref{elastic-Gamma}) can be evaluated explicitly. As an example, we take a dot with even number of electrons $N$ and a ground-state spin $S_i=0$. The intermediate states $j$ in Eq.~(\ref{elastic-Gamma}) for the dot with $N\pm 1$ electrons have spin $S_j=1/2$ (in the case of $S_i \neq 0$, there are two possible values $S_j=S_i \pm 1/2$). Since orbital occupations are good quantum numbers for the universal Hamiltonian, the sums over $j$ reduce to sums over single-particle levels
\begin{equation}\label{elastic-Gamma-UH}
\Gamma_{ii}^{ab}=\frac{\hbar}{2 \pi} \Gamma_{0}^{a}(\epsilon) \Gamma_{0}^{b}(\epsilon) \left| \sum_{\epsilon_\lambda>0}\frac{\phi_{\lambda}({\bf r}_{r})\phi_{\lambda}^{*}({\bf r}_{l})}
{\epsilon_{\lambda}-\frac{3}{4} J_s - \tilde{\epsilon}_{F}-\epsilon} + \sum_{\epsilon_\lambda<0}
\frac{\phi_{\lambda}({\bf r}_{r})\phi_{\lambda}^{*}({\bf r}_{l})}
{-\epsilon_{\lambda} -\frac{3}{4} J_s  + \tilde{\epsilon}_{F} + E_c +\epsilon}\right|^{2} \;,
\end{equation}
where $\Gamma_{0}^{a}={2 \frac{\pi}{ \hbar}}  \sum_{k} \delta(\epsilon_k - \epsilon_a)|t_{k}^{a}|^{2}$, $\phi_{\lambda}({\bf r})$ is the orbital single-particle wave function $\lambda$, and $\epsilon_\lambda$ is measured with respect to the Fermi energy.

We next calculate the average of (\ref{elastic-Gamma-UH}) over the mesoscopic fluctuations.~\cite{ABG02} Assuming the wave function amplitudes at the left and right  point contacts are uncorrelated and using $\overline{\phi_{\lambda}({\bf r}_{a})\phi^*_{\mu}({\bf r}_{a})}=\delta_{\lambda \mu}/{\cal A}$ (${\cal A}$ is the area of the dot), we obtain
\begin{eqnarray}
\overline{\Gamma_{ii}^{lr}(\epsilon)}=\frac{\hbar}{2 \pi} \Gamma_{0}^l \Gamma_{0}^r \frac{1}{{\cal A}^2} \left[\left\langle \sum_{\epsilon_\lambda>0} \frac{1}{\left(\epsilon_\lambda - \frac{3}{4} J_s -\tilde{\epsilon}_{F} -\epsilon\right)^2 } \right\rangle + \left\langle \sum_{\epsilon_\lambda< 0}\frac{1}{\left(-\epsilon_\lambda - \frac{3}{4} J_s + \tilde{\epsilon}_{F}+ E_c +\epsilon\right)^2} \right\rangle \right] \;.
\end{eqnarray}
where the remaining average is over the single-particle level fluctuations. Replacing the sums over single-particle levels by integrals we obtain
\begin{equation}
\overline{\Gamma_{ii}^{lr}(\epsilon)}=\frac{\hbar}{2 \pi} \Gamma_{0}^l \Gamma_{0}^r \frac{1}{{\cal A}^{2}\delta} \left(\frac{1}{-\tilde{\epsilon}_{F} - \frac{3}{4} J_s -\epsilon}+\frac{1}{\tilde{\epsilon}_{F}+e^{2}/C - \frac{3}{4} J_s +\epsilon}\right) \;.
\label{gammabar}
\end{equation}

Using Eq.~(\ref{gammabar}) in expressions (\ref{elastic-low-T}), and taking $\tilde P_i^{(N)}\approx 1$ away from the degeneracy point, we obtain Eqs.~(\ref{average-G-co}) and (\ref{average-GT-co}) for the average values of the electrical and thermal conductances in elastic cotunneling.

\bibliography{gbbib}

%Merlin.mbs v4.21 2009-07-09.
\begin{thebibliography}{10}%
\makeatletter
\providecommand \@ifxundefined [1]{%
 \ifx #1\undefined \expandafter \@firstoftwo
 \else \expandafter \@secondoftwo
\fi
}%
\providecommand \@ifnum [1]{%
 \ifnum #1\expandafter \@firstoftwo
 \else \expandafter \@secondoftwo
\fi
}%
\providecommand \enquote [1]{``#1''}%
\providecommand \bibnamefont  [1]{#1}%
\providecommand \bibfnamefont [1]{#1}%
\providecommand \citenamefont [1]{#1}%
\providecommand\href[0]{\@sanitize\@href}%
\providecommand\@href[1]{\endgroup\@@startlink{#1}\endgroup\@@href}%
\providecommand\@@href[1]{#1\@@endlink}%
\providecommand \@sanitize [0]{\begingroup\catcode`\&12\catcode`\#12\relax}%
\@ifxundefined \pdfoutput {\@firstoftwo}{%
 \@ifnum{\z@=\pdfoutput}{\@firstoftwo}{\@secondoftwo}%
}{%
 \providecommand\@@startlink[1]{\leavevmode\special{html:<a href="#1">}}%
 \providecommand\@@endlink[0]{\special{html:</a>}}%
}{%
 \providecommand\@@startlink[1]{%
  \leavevmode
  \pdfstartlink
   attr{/Border[0 0 1 ]/H/I/C[0 1 1]}%
   user{/Subtype/Link/A<</Type/Action/S/URI/URI(#1)>>}%
  \relax
 }%
 \providecommand\@@endlink[0]{\pdfendlink}%
}%
\providecommand \url  [0]{\begingroup\@sanitize \@url }%
\providecommand \@url [1]{\endgroup\@href {#1}{\urlprefix}}%
\providecommand \urlprefix [0]{URL }%
\providecommand \Eprint[0]{\href }%
\@ifxundefined \urlstyle {%
  \providecommand \doi [1]{doi:\discretionary{}{}{}#1}%
}{%
  \providecommand \doi [0]{doi:\discretionary{}{}{}\begingroup
  \urlstyle{rm}\Url }%
}%
\providecommand \doibase [0]{http://dx.doi.org/}%
\providecommand \Doi[1]{\href{\doibase#1}}%
\providecommand \bibAnnote [3]{%
  \BibitemShut{#1}%
  \begin{quotation}\noindent
    \textsc{Key:}\ #2\\\textsc{Annotation:}\ #3%
  \end{quotation}%
}%
\providecommand \bibAnnoteFile [2]{%
  \IfFileExists{#2}{\bibAnnote {#1} {#2} {\input{#2}}}{}%
}%
\providecommand \typeout [0]{\immediate \write \m@ne }%
\providecommand \selectlanguage [0]{\@gobble}%
\providecommand \bibinfo [0]{\@secondoftwo}%
\providecommand \bibfield [0]{\@secondoftwo}%
\providecommand \translation [1]{[#1]}%
\providecommand \BibitemOpen[0]{}%
\providecommand \bibitemStop [0]{}%
\providecommand \bibitemNoStop [0]{.\EOS\space}%
\providecommand \EOS [0]{\spacefactor3000\relax}%
\providecommand \BibitemShut [1]{\csname bibitem#1\endcsname}%
%</preamble>
\bibitem{marcusreview}%
  \BibitemOpen
  \bibfield{author}{%
  \bibinfo {author} {\bibfnamefont{L.~P.}\ \bibnamefont{Kouwenhoven}}\ and\
  \bibinfo {author} {\bibfnamefont{C.~M.}\ \bibnamefont{Marcus}},\ }%
  \bibfield{journal}{%
  \bibinfo {journal} {Physics World}}%
   (\bibinfo {month} {June}\ \bibinfo {year} {1998})%
  \bibAnnoteFile{NoStop}{marcusreview}%
\bibitem{alhassid2}%
  \BibitemOpen
  \bibfield{author}{%
  \bibinfo {author} {\bibfnamefont{Y.}~\bibnamefont{Alhassid}},\ }%
  \bibfield{journal}{%
  \bibinfo {journal} {Rev. Mod. Phys}\ }%
  \textbf{\bibinfo {volume} {72}},\ \bibinfo {pages} {895} (\bibinfo {year}
  {2000})%
  \bibAnnoteFile{NoStop}{alhassid2}%
\bibitem{JSA}%
  \BibitemOpen
  \bibfield{author}{%
  \bibinfo {author} {\bibfnamefont{R.~A.}\ \bibnamefont{Jalabert}}, \bibinfo
  {author} {\bibfnamefont{A.~D.}\ \bibnamefont{Stone}},\ and\ \bibinfo {author}
  {\bibfnamefont{Y.}~\bibnamefont{Alhassid}},\ }%
  \bibfield{journal}{%
  \bibinfo {journal} {Phys. Rev. Lett.}\ }%
  \textbf{\bibinfo {volume} {68}},\ \bibinfo {pages} {3468} (\bibinfo {year}
  {1992})%
  \bibAnnoteFile{NoStop}{JSA}%
\bibitem{folk}%
  \BibitemOpen
  \bibfield{author}{%
  \bibinfo {author} {\bibfnamefont{J.~A.}\ \bibnamefont{Folk}}, \bibinfo
  {author} {\bibfnamefont{S.~R.}\ \bibnamefont{Patel}}, \bibinfo {author}
  {\bibfnamefont{S.~F.}\ \bibnamefont{Godijn}}, \bibinfo {author}
  {\bibfnamefont{A.~G.}\ \bibnamefont{Huibers}}, \bibinfo {author}
  {\bibfnamefont{S.~M.}\ \bibnamefont{Cronenwett}}, \bibinfo {author}
  {\bibfnamefont{C.~M.}\ \bibnamefont{Marcus}}, \bibinfo {author}
  {\bibfnamefont{K.}~\bibnamefont{Campman}},\ and\ \bibinfo {author}
  {\bibfnamefont{A.~C.}\ \bibnamefont{Gossard}},\ }%
  \bibfield{journal}{%
  \bibinfo {journal} {Phys. Rev. Lett.}\ }%
  \textbf{\bibinfo {volume} {76}},\ \bibinfo {pages} {1699} (\bibinfo {year}
  {1996})%
  \bibAnnoteFile{NoStop}{folk}%
\bibitem{chang}%
  \BibitemOpen
  \bibfield{author}{%
  \bibinfo {author} {\bibfnamefont{A.~M.}\ \bibnamefont{Chang}}, \bibinfo
  {author} {\bibfnamefont{H.~U.}\ \bibnamefont{Baranger}}, \bibinfo {author}
  {\bibfnamefont{L.~N.}\ \bibnamefont{Pfeiffer}}, \bibinfo {author}
  {\bibfnamefont{K.~W.}\ \bibnamefont{West}},\ and\ \bibinfo {author}
  {\bibfnamefont{T.~Y.}\ \bibnamefont{Chang}},\ }%
  \bibfield{journal}{%
  \bibinfo {journal} {Phys. Rev. Lett.}\ }%
  \textbf{\bibinfo {volume} {76}},\ \bibinfo {pages} {1695} (\bibinfo {year}
  {1996})%
  \bibAnnoteFile{NoStop}{chang}%
\bibitem{sivan96}%
  \BibitemOpen
  \bibfield{author}{%
  \bibinfo {author} {\bibfnamefont{U.}~\bibnamefont{Sivan}}, \bibinfo {author}
  {\bibfnamefont{R.}~\bibnamefont{Berkovits}}, \bibinfo {author}
  {\bibfnamefont{Y.}~\bibnamefont{Aloni}}, \bibinfo {author}
  {\bibfnamefont{O.}~\bibnamefont{Prus}}, \bibinfo {author}
  {\bibfnamefont{A.}~\bibnamefont{Auerbach}},\ and\ \bibinfo {author}
  {\bibfnamefont{G.}~\bibnamefont{Ben-Yoseph}},\ }%
  \bibfield{journal}{%
  \bibinfo {journal} {Phys. Rev. Lett.}\ }%
  \textbf{\bibinfo {volume} {77}},\ \bibinfo {pages} {1123} (\bibinfo {year}
  {1996})%
  \bibAnnoteFile{NoStop}{sivan96}%
\bibitem{univH}%
  \BibitemOpen
  \bibfield{author}{%
  \bibinfo {author} {\bibfnamefont{I.~L.}\ \bibnamefont{Kurland}}, \bibinfo
  {author} {\bibfnamefont{I.~L.}\ \bibnamefont{Aleiner}},\ and\ \bibinfo
  {author} {\bibfnamefont{B.~L.}\ \bibnamefont{Altshuler}},\ }%
  \bibfield{journal}{%
  \bibinfo {journal} {Phys. Rev. B}\ }%
  \textbf{\bibinfo {volume} {62}},\ \bibinfo {pages} {14886} (\bibinfo {year}
  {2000})%
  \bibAnnoteFile{NoStop}{univH}%
\bibitem{ABG02}%
  \BibitemOpen
  \bibfield{author}{%
  \bibinfo {author} {\bibfnamefont{I.}~\bibnamefont{Aleiner}}, \bibinfo
  {author} {\bibfnamefont{P.}~\bibnamefont{Brouwer}},\ and\ \bibinfo {author}
  {\bibfnamefont{L.}~\bibnamefont{Glazman}},\ }%
  \bibfield{journal}{%
  \bibinfo {journal} {Phys. Rep.}\ }%
  \textbf{\bibinfo {volume} {358}},\ \bibinfo {pages} {309} (\bibinfo {year}
  {2002})%
  \bibAnnoteFile{NoStop}{ABG02}%
\bibitem{alhassid3}%
  \BibitemOpen
  \bibfield{author}{%
  \bibinfo {author} {\bibfnamefont{Y.}~\bibnamefont{Alhassid}}\ and\ \bibinfo
  {author} {\bibfnamefont{T.}~\bibnamefont{Rupp}},\ }%
  \bibfield{journal}{%
  \bibinfo {journal} {Phys. Rev. Lett.}\ }%
  \textbf{\bibinfo {volume} {91}},\ \bibinfo {pages} {056801} (\bibinfo {year}
  {2003})%
  \bibAnnoteFile{NoStop}{alhassid3}%
\bibitem{patel1}%
  \BibitemOpen
  \bibfield{author}{%
  \bibinfo {author} {\bibfnamefont{S.~R.}\ \bibnamefont{Patel}}, \bibinfo
  {author} {\bibfnamefont{S.~M.}\ \bibnamefont{Cronenwett}}, \bibinfo {author}
  {\bibfnamefont{D.~R.}\ \bibnamefont{Stewart}}, \bibinfo {author}
  {\bibfnamefont{A.~G.}\ \bibnamefont{Huibers}}, \bibinfo {author}
  {\bibfnamefont{C.~M.}\ \bibnamefont{Marcus}}, \bibinfo {author}
  {\bibfnamefont{C.~I.}\ \bibnamefont{Duru{\"{o}}z}}, \bibinfo {author}
  {\bibfnamefont{J.~S.}\ \bibnamefont{Harris}}, \bibinfo {author}
  {\bibfnamefont{K.}~\bibnamefont{Campman}},\ and\ \bibinfo {author}
  {\bibfnamefont{A.~C.}\ \bibnamefont{Gossard}},\ }%
  \bibfield{journal}{%
  \bibinfo {journal} {Phys. Rev. Lett.}\ }%
  \textbf{\bibinfo {volume} {80}},\ \bibinfo {pages} {4522} (\bibinfo {year}
  {1998})%
  \bibAnnoteFile{NoStop}{patel1}%
\bibitem{patel2}%
  \BibitemOpen
  \bibfield{author}{%
  \bibinfo {author} {\bibfnamefont{S.~R.}\ \bibnamefont{Patel}}, \bibinfo
  {author} {\bibfnamefont{D.~R.}\ \bibnamefont{Stewart}}, \bibinfo {author}
  {\bibfnamefont{C.~M.}\ \bibnamefont{Marcus}}, \bibinfo {author}
  {\bibfnamefont{M.}~\bibnamefont{G{\"o}k{\c c}eda{\u g}}}, \bibinfo {author}
  {\bibfnamefont{Y.}~\bibnamefont{Alhassid}}, \bibinfo {author}
  {\bibfnamefont{A.~D.}\ \bibnamefont{Stone}}, \bibinfo {author}
  {\bibfnamefont{C.~I.}\ \bibnamefont{Duru{\"o}z}},\ and\ \bibinfo {author}
  {\bibfnamefont{J.~S.}\ \bibnamefont{Harris}},\ }%
  \bibfield{journal}{%
  \bibinfo {journal} {Phys. Rev. Lett.}\ }%
  \textbf{\bibinfo {volume} {81}},\ \bibinfo {pages} {5900} (\bibinfo {year}
  {1998})%
  \bibAnnoteFile{NoStop}{patel2}%
\bibitem{beenakker1}%
  \BibitemOpen
  \bibfield{author}{%
  \bibinfo {author} {\bibfnamefont{C.~W.~J.}\ \bibnamefont{Beenakker}}\ and\
  \bibinfo {author} {\bibfnamefont{A.~A.~M.}\ \bibnamefont{Staring}},\ }%
  \bibfield{journal}{%
  \bibinfo {journal} {Phys. Rev. B}\ }%
  \textbf{\bibinfo {volume} {46}},\ \bibinfo {pages} {9667} (\bibinfo {year}
  {1992})%
  \bibAnnoteFile{NoStop}{beenakker1}%
\bibitem{staring}%
  \BibitemOpen
  \bibfield{author}{%
  \bibinfo {author} {\bibfnamefont{A.~A.~M.}\ \bibnamefont{Staring}}, \bibinfo
  {author} {\bibfnamefont{L.~W.}\ \bibnamefont{Molenkamp}}, \bibinfo {author}
  {\bibfnamefont{B.~W.}\ \bibnamefont{Alphenaar}}, \bibinfo {author}
  {\bibfnamefont{H.~V.}\ \bibnamefont{Houten}}, \bibinfo {author}
  {\bibfnamefont{O.~J.~A.}\ \bibnamefont{Buyk}}, \bibinfo {author}
  {\bibfnamefont{M.~A.~A.}\ \bibnamefont{Mabesoone}}, \bibinfo {author}
  {\bibfnamefont{C.~W.~J.}\ \bibnamefont{Beenakker}},\ and\ \bibinfo {author}
  {\bibfnamefont{C.~T.}\ \bibnamefont{Foxon}},\ }%
  \bibfield{journal}{%
  \bibinfo {journal} {Europhys. Lett.}\ }%
  \textbf{\bibinfo {volume} {22}},\ \bibinfo {pages} {57} (\bibinfo {year}
  {1993})%
  \bibAnnoteFile{NoStop}{staring}%
\bibitem{dzurak1}%
  \BibitemOpen
  \bibfield{author}{%
  \bibinfo {author} {\bibfnamefont{A.~S.}\ \bibnamefont{Dzurak}}, \bibinfo
  {author} {\bibfnamefont{C.~G.}\ \bibnamefont{Smith}}, \bibinfo {author}
  {\bibfnamefont{D.~A.}\ \bibnamefont{Ritchie}}, \bibinfo {author}
  {\bibfnamefont{J.~E.~F.}\ \bibnamefont{Frost}}, \bibinfo {author}
  {\bibfnamefont{G.~A.~C.}\ \bibnamefont{Jones}},\ and\ \bibinfo {author}
  {\bibfnamefont{D.~G.}\ \bibnamefont{Hasko}},\ }%
  \bibfield{journal}{%
  \bibinfo {journal} {Solid State Commun.}\ }%
  \textbf{\bibinfo {volume} {87}},\ \bibinfo {pages} {1145} (\bibinfo {year}
  {1993})%
  \bibAnnoteFile{NoStop}{dzurak1}%
\bibitem{dzurak2}%
  \BibitemOpen
  \bibfield{author}{%
  \bibinfo {author} {\bibfnamefont{A.~S.}\ \bibnamefont{Dzurak}}, \bibinfo
  {author} {\bibfnamefont{C.~G.}\ \bibnamefont{Smith}}, \bibinfo {author}
  {\bibfnamefont{C.~H.~W.}\ \bibnamefont{Barnes}}, \bibinfo {author}
  {\bibfnamefont{M.}~\bibnamefont{Pepper}}, \bibinfo {author}
  {\bibfnamefont{L.}~\bibnamefont{Mart\'{i}n-Moreno}}, \bibinfo {author}
  {\bibfnamefont{C.~T.}\ \bibnamefont{Liang}}, \bibinfo {author}
  {\bibfnamefont{D.~A.}\ \bibnamefont{Ritchie}},\ and\ \bibinfo {author}
  {\bibfnamefont{G.~A.~C.}\ \bibnamefont{Jones}},\ }%
  \bibfield{journal}{%
  \bibinfo {journal} {Phys. Rev. B}\ }%
  \textbf{\bibinfo {volume} {55}},\ \bibinfo {pages} {R 10197} (\bibinfo {year}
  {1997})%
  \bibAnnoteFile{NoStop}{dzurak2}%
\bibitem{turek}%
  \BibitemOpen
  \bibfield{author}{%
  \bibinfo {author} {\bibfnamefont{M.}~\bibnamefont{Turek}}\ and\ \bibinfo
  {author} {\bibfnamefont{K.~A.}\ \bibnamefont{Matveev}},\ }%
  \bibfield{journal}{%
  \bibinfo {journal} {Phys Rev. B}\ }%
  \textbf{\bibinfo {volume} {65}},\ \bibinfo {pages} {115332} (\bibinfo {year}
  {2002})%
  \bibAnnoteFile{NoStop}{turek}%
\bibitem{koch04}%
  \BibitemOpen
  \bibfield{author}{%
  \bibinfo {author} {\bibfnamefont{J.}~\bibnamefont{Koch}}, \bibinfo {author}
  {\bibfnamefont{F.}~\bibnamefont{von Oppen}}, \bibinfo {author}
  {\bibfnamefont{Y.}~\bibnamefont{Oreg}},\ and\ \bibinfo {author}
  {\bibfnamefont{E.}~\bibnamefont{Sela}},\ }%
  \bibfield{journal}{%
  \bibinfo {journal} {Phys. Rev. B}\ }%
  \textbf{\bibinfo {volume} {70}},\ \bibinfo {pages} {195107} (\bibinfo {year}
  {2004})%
  \bibAnnoteFile{NoStop}{koch04}%
\bibitem{beenakker2}%
  \BibitemOpen
  \bibfield{author}{%
  \bibinfo {author} {\bibfnamefont{C.~W.~J.}\ \bibnamefont{Beenakker}},\ }%
  \bibfield{journal}{%
  \bibinfo {journal} {Phys. Rev. B}\ }%
  \textbf{\bibinfo {volume} {44}},\ \bibinfo {pages} {1646} (\bibinfo {year}
  {1991})%
  \bibAnnoteFile{NoStop}{beenakker2}%
\bibitem{alhassid1}%
  \BibitemOpen
  \bibfield{author}{%
  \bibinfo {author} {\bibfnamefont{Y.}~\bibnamefont{Alhassid}}, \bibinfo
  {author} {\bibfnamefont{T.}~\bibnamefont{Rupp}}, \bibinfo {author}
  {\bibfnamefont{A.}~\bibnamefont{Kaminski}},\ and\ \bibinfo {author}
  {\bibfnamefont{L.~I.}\ \bibnamefont{Glazman}},\ }%
  \bibfield{journal}{%
  \bibinfo {journal} {Phys. Rev. B}\ }%
  \textbf{\bibinfo {volume} {69}},\ \bibinfo {pages} {115331} (\bibinfo {year}
  {2004})%
  \bibAnnoteFile{NoStop}{alhassid1}%
\bibitem{brouwer99}%
  \BibitemOpen
  \bibfield{author}{%
  \bibinfo {author} {\bibfnamefont{P.~W.}\ \bibnamefont{Brouwer}}, \bibinfo
  {author} {\bibfnamefont{Y.}~\bibnamefont{Oreg}},\ and\ \bibinfo {author}
  {\bibfnamefont{B.~I.}\ \bibnamefont{Halperin}},\ }%
  \bibfield{journal}{%
  \bibinfo {journal} {Phys. Rev. B}\ }%
  \textbf{\bibinfo {volume} {60}},\ \bibinfo {pages} {R 13977} (\bibinfo {year}
  {1999})%
  \bibAnnoteFile{NoStop}{brouwer99}%
\bibitem{baranger00}%
  \BibitemOpen
  \bibfield{author}{%
  \bibinfo {author} {\bibfnamefont{H.~U.}\ \bibnamefont{Baranger}}, \bibinfo
  {author} {\bibfnamefont{D.}~\bibnamefont{Ullmo}},\ and\ \bibinfo {author}
  {\bibfnamefont{L.~I.}\ \bibnamefont{Glazman}},\ }%
  \bibfield{journal}{%
  \bibinfo {journal} {Phys. Rev. B}\ }%
  \textbf{\bibinfo {volume} {61}},\ \bibinfo {pages} {R2425} (\bibinfo {year}
  {2000})%
  \bibAnnoteFile{NoStop}{baranger00}%
\bibitem{alhassid02}%
  \BibitemOpen
  \bibfield{author}{%
  \bibinfo {author} {\bibfnamefont{Y.}~\bibnamefont{Alhassid}}\ and\ \bibinfo
  {author} {\bibfnamefont{S.}~\bibnamefont{Malhotra}},\ }%
  \bibfield{journal}{%
  \bibinfo {journal} {Phys. Rev. B}\ }%
  \textbf{\bibinfo {volume} {66}},\ \bibinfo {pages} {245313} (\bibinfo {year}
  {2002})%
  \bibAnnoteFile{NoStop}{alhassid02}%
\bibitem{kiselev}%
  \BibitemOpen
  \bibfield{author}{%
  \bibinfo {author} {\bibfnamefont{M.~N.}\ \bibnamefont{Kiselev}}\ and\
  \bibinfo {author} {\bibfnamefont{Y.}~\bibnamefont{Gefen}},\ }%
  \bibfield{journal}{%
  \bibinfo {journal} {Phys. Rev. Lett.}\ }%
  \textbf{\bibinfo {volume} {96}},\ \bibinfo {pages} {066805} (\bibinfo {year}
  {2006})%
  \bibAnnoteFile{NoStop}{kiselev}%
\bibitem{kiselev2}%
  \BibitemOpen
  \bibfield{author}{%
  \bibinfo {author} {\bibfnamefont{I.~S.}\ \bibnamefont{Burmistrov}}, \bibinfo
  {author} {\bibfnamefont{Y.}~\bibnamefont{Gefen}},\ and\ \bibinfo {author}
  {\bibfnamefont{M.~N.}\ \bibnamefont{Kiselev}}}%
   (\bibinfo {year} {2009}),\
  \Eprint{http://arxiv.org/abs/0912.3185}{arXiv:0912.3185}%
  \bibAnnoteFile{NoStop}{kiselev2}%
\bibitem{kondo}%
  \BibitemOpen
  \bibfield{author}{%
  \bibinfo {author} {\bibfnamefont{D.}~\bibnamefont{Boese}}\ and\ \bibinfo
  {author} {\bibfnamefont{R.}~\bibnamefont{Fazio}},\ }%
  \bibfield{journal}{%
  \bibinfo {journal} {Europhys. Lett.}\ }%
  \textbf{\bibinfo {volume} {56}},\ \bibinfo {pages} {576} (\bibinfo {year}
  {2001})%
  \bibAnnoteFile{NoStop}{kondo}%
\bibitem{nguyen09}%
  \BibitemOpen
  \bibfield{author}{%
  \bibinfo {author} {\bibfnamefont{T.~K.~T.}\ \bibnamefont{Nguyen}}, \bibinfo
  {author} {\bibfnamefont{M.~N.}\ \bibnamefont{Kiselev}},\ and\ \bibinfo
  {author} {\bibfnamefont{V.~E.}\ \bibnamefont{Kravtsov}}}%
   (\bibinfo {year} {2009}),\
  \Eprint{http://arxiv.org/abs/0912.4632}{arXiv:0912.4632}%
  \bibAnnoteFile{NoStop}{nguyen09}%
\bibitem{Note1}%
  \BibitemOpen
  \bibinfo {note} {For a dot with left-right symmetry (i.e., $\Gamma _{ij}^l =
  \Gamma _{ij}^r$ for all $i,j$), Eq.~(\ref {thermopowereq}) is exact for a
  general interaction in the dot (see Appendix A).}%
  \bibAnnoteFile{Stop}{Note1}%
\bibitem{Note2}%
  \BibitemOpen
  \bibinfo {note} {When the competing cotunneling process is included this is
  no longer the case, and the smallness of $\protect \mathaccentV
  {tilde}07E{P}_{0}^{(N+1)}$ (or $\protect \mathaccentV
  {tilde}07E{P}_{0}^{(N)}$) does reduce the thermopower (see Sec.~\ref
  {cotunneling}).}%
  \bibAnnoteFile{Stop}{Note2}%
\bibitem{Note3}%
  \BibitemOpen
  \bibinfo {note} {We have assumed here energy-independent transition widths.
  For realistic dots, one has to average over the mesoscopic fluctuations to
  recover this limit.}%
  \bibAnnoteFile{Stop}{Note3}%
\bibitem{parallel1}%
  \BibitemOpen
  \bibfield{author}{%
  \bibinfo {author} {\bibfnamefont{D.~S.}\ \bibnamefont{Duncan}}, \bibinfo
  {author} {\bibfnamefont{D.}~\bibnamefont{Goldhaber-Gordon}}, \bibinfo
  {author} {\bibfnamefont{R.~M.}\ \bibnamefont{Westervelt}}, \bibinfo {author}
  {\bibfnamefont{K.~D.}\ \bibnamefont{Maranowski}},\ and\ \bibinfo {author}
  {\bibfnamefont{A.~C.}\ \bibnamefont{Gossard}},\ }%
  \bibfield{journal}{%
  \bibinfo {journal} {Appl.~Phys.~Lett.~}\ }%
  \textbf{\bibinfo {volume} {77}},\ \bibinfo {pages} {2183} (\bibinfo {year}
  {2000})%
  \bibAnnoteFile{NoStop}{parallel1}%
\bibitem{parallel2}%
  \BibitemOpen
  \bibfield{author}{%
  \bibinfo {author} {\bibfnamefont{S.}~\bibnamefont{Lindemann}}, \bibinfo
  {author} {\bibfnamefont{T.}~\bibnamefont{Ihn}}, \bibinfo {author}
  {\bibfnamefont{T.}~\bibnamefont{Heinzel}}, \bibinfo {author}
  {\bibfnamefont{W.}~\bibnamefont{Zwerger}}, \bibinfo {author}
  {\bibfnamefont{K.}~\bibnamefont{Ensslin}}, \bibinfo {author}
  {\bibfnamefont{K.}~\bibnamefont{Maranowski}},\ and\ \bibinfo {author}
  {\bibfnamefont{A.~C.}\ \bibnamefont{Gossard}},\ }%
  \bibfield{journal}{%
  \bibinfo {journal} {Phys. Rev. B}\ }%
  \textbf{\bibinfo {volume} {66}},\ \bibinfo {pages} {195314} (\bibinfo {year}
  {2002})%
  \bibAnnoteFile{NoStop}{parallel2}%
\bibitem{parallel3}%
  \BibitemOpen
  \bibfield{author}{%
  \bibinfo {author} {\bibfnamefont{R.~M.}\ \bibnamefont{Potok}}, \bibinfo
  {author} {\bibfnamefont{J.~A.}\ \bibnamefont{Folk}}, \bibinfo {author}
  {\bibfnamefont{C.~M.}\ \bibnamefont{Marcus}}, \bibinfo {author}
  {\bibfnamefont{V.}~\bibnamefont{Umansky}}, \bibinfo {author}
  {\bibfnamefont{M.}~\bibnamefont{Hanson}},\ and\ \bibinfo {author}
  {\bibfnamefont{A.~C.}\ \bibnamefont{Gossard}},\ }%
  \bibfield{journal}{%
  \bibinfo {journal} {Phys. Rev. Lett.}\ }%
  \textbf{\bibinfo {volume} {91}},\ \bibinfo {pages} {016802} (\bibinfo {year}
  {2003})%
  \bibAnnoteFile{NoStop}{parallel3}%
\bibitem{huertas}%
  \BibitemOpen
  \bibfield{author}{%
  \bibinfo {author} {\bibfnamefont{D.}~\bibnamefont{Huertas-Hernando}}\ and\
  \bibinfo {author} {\bibfnamefont{Y.}~\bibnamefont{Alhassid}},\ }%
  \bibfield{journal}{%
  \bibinfo {journal} {Phys. Rev. B}\ }%
  \textbf{\bibinfo {volume} {75}},\ \bibinfo {pages} {153312} (\bibinfo {year}
  {2007})%
  \bibAnnoteFile{NoStop}{huertas}%
\bibitem{ensslin08}%
  \BibitemOpen
  \bibfield{author}{%
  \bibinfo {author} {\bibfnamefont{S.}~\bibnamefont{Gustavsson}}, \bibinfo
  {author} {\bibfnamefont{R.}~\bibnamefont{Leturcq}}, \bibinfo {author}
  {\bibfnamefont{M.}~\bibnamefont{Studer}}, \bibinfo {author}
  {\bibfnamefont{T.}~\bibnamefont{Ihn}}, \bibinfo {author}
  {\bibfnamefont{K.}~\bibnamefont{Ensslin}}, \bibinfo {author}
  {\bibfnamefont{D.~C.}\ \bibnamefont{Driscoll}},\ and\ \bibinfo {author}
  {\bibfnamefont{A.~C.}\ \bibnamefont{Gossard}},\ }%
  \bibfield{journal}{%
  \bibinfo {journal} {Nano Letters}\ }%
  \textbf{\bibinfo {volume} {8}},\ \bibinfo {pages} {2547} (\bibinfo {year}
  {2008})%
  \bibAnnoteFile{NoStop}{ensslin08}%
\bibitem{cronenwett}%
  \BibitemOpen
  \bibfield{author}{%
  \bibinfo {author} {\bibfnamefont{S.~M.}\ \bibnamefont{Cronenwett}}, \bibinfo
  {author} {\bibfnamefont{S.~R.}\ \bibnamefont{Patel}}, \bibinfo {author}
  {\bibfnamefont{C.~M.}\ \bibnamefont{Marcus}}, \bibinfo {author}
  {\bibfnamefont{K.}~\bibnamefont{Campman}},\ and\ \bibinfo {author}
  {\bibfnamefont{A.~C.}\ \bibnamefont{Gossard}},\ }%
  \bibfield{journal}{%
  \bibinfo {journal} {Phys. Rev. Lett.}\ }%
  \textbf{\bibinfo {volume} {79}},\ \bibinfo {pages} {2312} (\bibinfo {year}
  {1997})%
  \bibAnnoteFile{NoStop}{cronenwett}%
\bibitem{averin}%
  \BibitemOpen
  \bibfield{author}{%
  \bibinfo {author} {\bibfnamefont{D.~V.}\ \bibnamefont{Averin}}\ and\ \bibinfo
  {author} {\bibfnamefont{Y.~V.}\ \bibnamefont{Nazarov}},\ }%
  \bibfield{journal}{%
  \bibinfo {journal} {Phys. Rev. Lett.}\ }%
  \textbf{\bibinfo {volume} {65}},\ \bibinfo {pages} {2446} (\bibinfo {year}
  {1990})%
  \bibAnnoteFile{NoStop}{averin}%
\bibitem{aleiner}%
  \BibitemOpen
  \bibfield{author}{%
  \bibinfo {author} {\bibfnamefont{I.~L.}\ \bibnamefont{Aleiner}}\ and\
  \bibinfo {author} {\bibfnamefont{L.~I.}\ \bibnamefont{Glazman}},\ }%
  \bibfield{journal}{%
  \bibinfo {journal} {Phys. Rev. Lett.}\ }%
  \textbf{\bibinfo {volume} {77}},\ \bibinfo {pages} {2057} (\bibinfo {year}
  {1996})%
  \bibAnnoteFile{NoStop}{aleiner}%
\bibitem{glazman02}%
  \BibitemOpen
  \bibfield{author}{%
  \bibinfo {author} {\bibfnamefont{L.}~\bibnamefont{Glazman}}\ and\ \bibinfo
  {author} {\bibfnamefont{M.}~\bibnamefont{Pustilnik}},\ }%
  in\ \emph{\bibinfo {booktitle} {New Directions in Mesoscopic Physics(Towards
  Nanoscience)}},\ \bibinfo {editor} {edited by\ \bibinfo {editor}
  {\bibfnamefont{R.}~\bibnamefont{Fazio}}, \bibinfo {editor}
  {\bibfnamefont{V.}~\bibnamefont{Gantmakher}},\ and\ \bibinfo {editor}
  {\bibfnamefont{Y.}~\bibnamefont{Imry}}}\ (\bibinfo {publisher} {Kluwer},\
  \bibinfo {address} {Dordrecht},\ \bibinfo {year} {2003})\ pp.\ \bibinfo
  {pages} {93--115}%
  \bibAnnoteFile{NoStop}{glazman02}%
\bibitem{ziman72}%
  \BibitemOpen
  \bibfield{author}{%
  \bibinfo {author} {\bibfnamefont{J.~M.}\ \bibnamefont{Ziman}},\ }%
  \emph{\bibinfo {title} {Principles of the Theory of Solids}}\ (\bibinfo
  {publisher} {Cambridge University Press},\ \bibinfo {address} {London},\
  \bibinfo {year} {1972})%
  \bibAnnoteFile{NoStop}{ziman72}%
\end{thebibliography}%

\end{document}